\documentstyle[psfig,general_cite]{mn}

\newcommand\etal{et al.}

\title{The Swinburne Intermediate Latitude Pulsar Survey}

\author[R.T. Edwards, M. Bailes, W. van Straten and M.C. Britton]
{R.T. Edwards, M. Bailes, W. van Straten and M.C. Britton\\
Centre for Astrophysics and Supercomputing, Swinburne
University of Technology, P.O.~Box 218 Hawthorn, VIC 3122, Australia}

\begin{document}
\maketitle

\begin{abstract}
We have conducted a survey of intermediate Galactic latitudes using
the 13-beam 21-cm multibeam receiver of the Parkes 64-m radio
telescope. The survey covered the region enclosed by $5\degr < |b| <
15\degr$ and $-100\degr < l < 50\degr$ with 4,702 processed pointings
of 265~s each, for a total of 14.5~days of integration time. Thirteen
$2\times96$-channel filterbanks provided 288 MHz of bandwidth at a
centre frequency of 1374~MHz, one-bit sampled every 125~$\mu$s and
incurring $\sim$DM$/13.4$~cm$^{-3}$~pc samples of dispersion
smearing. The system was sensitive to slow and most millisecond
pulsars in the region with flux densities greater than approximately
$0.3$--$1.1$~mJy. Offline analysis on the 64-node Swinburne
workstation cluster resulted in the detection of 170 pulsars of which
69 were new discoveries.  Eight of the new pulsars, by virtue of their
small spin periods and period derivatives, may be recycled and
have been reported elsewhere. The slow pulsars discovered are typical
of those already known in the volume searched, being of intermediate
to old age. Several pulsars experience pulse nulling and two display
very regular drifting sub-pulses.  We discuss the new discoveries and
provide timing parameters for the 48 slow pulsars for which we have a
phase-connnected solution.
\end{abstract}


\section{Introduction}
By the late 1990s radio pulsar surveys had resulted in the discovery
of $\sim$700 pulsars, spawning numerous studies with wide ranging
implications for astrophysics and physics in general. Despite having
been first discovered over a quarter of a century earlier, pulsars
with unique and interesting properties
(e.g. \pcite{wf92,jml+92,jlh+93,bbs+95,sbl+96}) continued to be
uncovered by surveys which also served the purpose of providing a
larger sample for statistical analyses of classes of pulsars and
pulsar binaries (e.g. \pcite{lml+98}).

Nearly all early surveys were conducted at low frequencies ($\nu
\simeq 400$~MHz) due to the steep spectrum ($\alpha \simeq -1.6$, where
$S\propto\nu^\alpha$; \pcite{lylg95}) characteristic of microwave radiation
from pulsars and the faster sky coverage afforded by the larger
telescope beam at these frequencies. However, two effects that hamper
the detection of certain pulsars at low frequencies can be avoided by
using a higher frequency. Firstly, for small Galactic latitudes the
background of Galactic synchrotron emission comprises the main
contribution to the system temperature at these frequencies. The
spectrum of this radiation is steep ($\alpha \simeq -2.6$;
\pcite{lmo+87}) and at high frequencies generally represents an
insignificant contribution compared to the thermal receiver
noise. Since they share the low Galactic $z$-height of their
progenitor population, young pulsars in particular are selected
against in low frequency surveys due to the elevated sky background
temperature. Secondly, radiation propagating through the interstellar
medium is subject to `scattering' due to multi-path propagation,
effectively convolving the light curve with an exponential of
a time constant that scales as $\nu^{-4}$ \cite{akh70}. Since the
minimum detectable mean flux density in pulsar observations is
proportional to $[\delta/(1-\delta)]^{1/2}$ where $\delta$ is the
effective pulse duty cycle, scatter-broadening of the received pulses
hampers the detection of pulsars at low frequencies, especially those
with short spin period such as the interesting and important class of
`millisecond' pulsars, and (again) young pulsars. 
Moreover, by conducting a survey at high frequencies one is
sensitive to pulsars with flatter spectra that were missed in earlier
surveys.

With the rise in availability of affordable computing power in the
1980s it became feasible to process surveys with fast sampling rates
and large numbers of pointings, as required for large scale
high frequency surveys for millisecond pulsars.
\scite{clj+92} and
\scite{jlm+92} conducted highly successful 20-cm pulsar surveys near
the Galactic plane, discovering 86 pulsars between them, including a
high fraction of young pulsars. However, the surveys did not have
sufficient sensitivity at high time resolution to discover any
millisecond pulsars. In addition, for the reasons mentioned above the
surveys concentrated on the Galactic plane and hence the samples
of detected pulsars were of reduced value in modelling the Galactic pulsar
population compared to larger surveys.

\begin{figure}
\centerline{\psfig{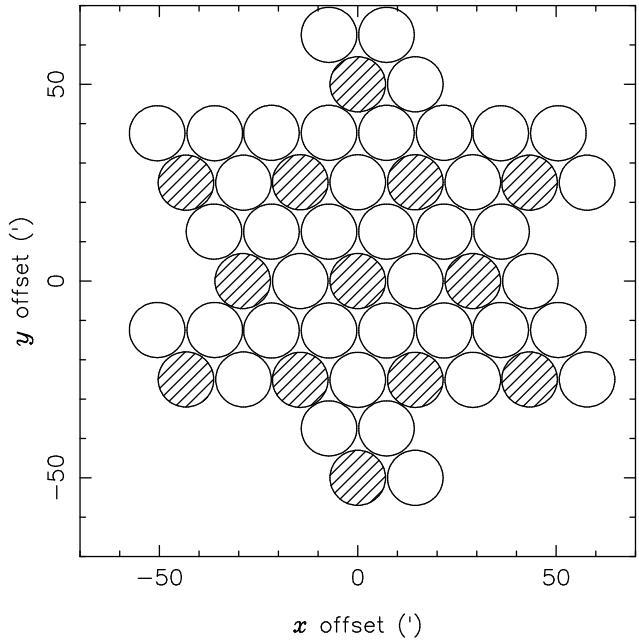}}
\caption{The multibeam tessellation unit shown with circles depicting
the half-power points of beams. A unit is observed with four offset
pointings, one of which is hatched in the above for clarity. The shape
made by the 52 beams can be seamlessly self-tessellated.}
\label{fig:mbshape}
\end{figure}
\begin{figure*}
\centerline{\psfig{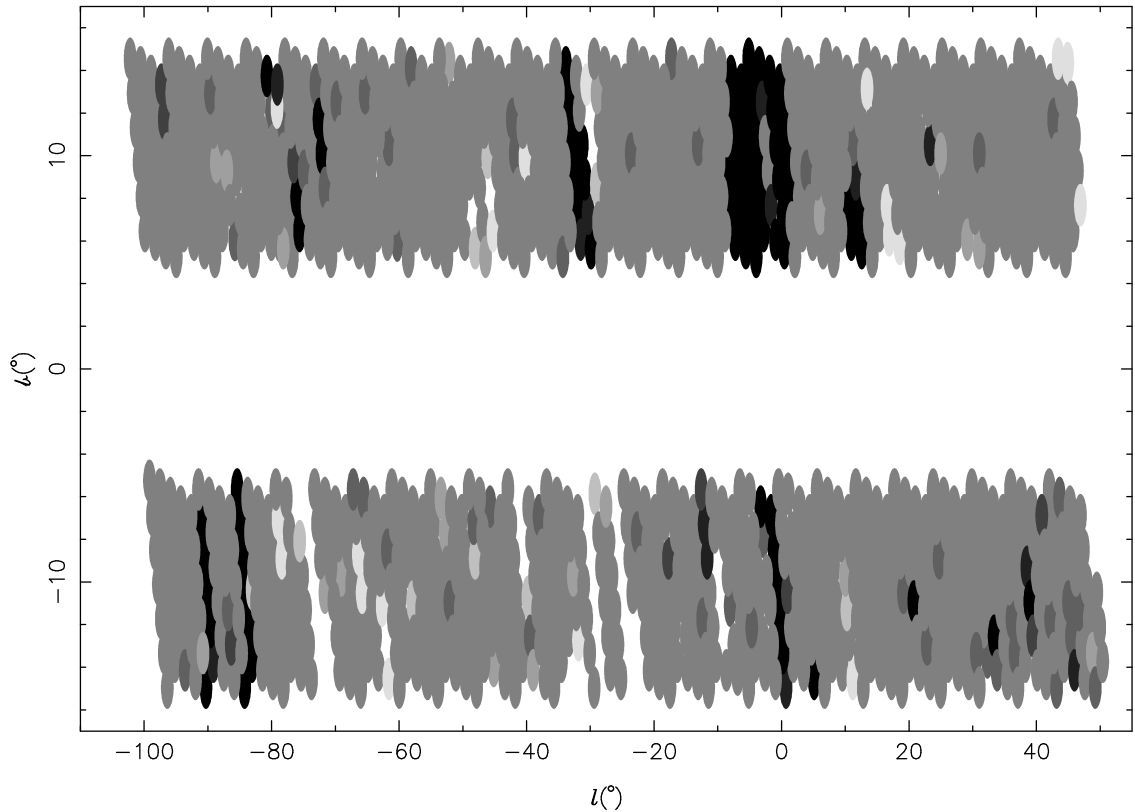}}
\caption{Sky coverage for the survey. Ellipses represent groups of four
inter-meshed pointings. Shades of grey represent the
density of coverage, from unobserved (white) to observed twice (black).}
\label{fig:coverage}
\end{figure*}

In 1997 the Australia Telescope National Facility commissioned a new
21-cm 13-feed multibeam receiver, primarily for HI surveys
\cite{hse+00,bsd+01}. The large instantaneous sky coverage and
excellent sensitivity also makes the system a powerful pulsar survey
instrument and this led to the commencement of a long-running deep
survey of the southern Galactic plane ($|b| < 5\degr$) which is
expected to almost double the known population
\cite{lcm+00,clm+00}. We conducted Monte Carlo simulations similar to
those discussed by \scite{tbms98} and found that a shallower
`flanking' survey should discover a sizeable population of pulsars
with unprecedented time efficiency in an area of sky not previously
sampled at high frequencies. Based on this result we conducted such a
survey between 1998 August and 1999 August. The survey proved highly
successful, discovering 69 pulsars including two pulsar binaries
containing heavy CO white dwarfs, one of which will coalesce in less
than a Hubble time with dramatic and unknown consequences
\cite{eb01a}, and a further four binary and two (perhaps three)
isolated recycled pulsars with important implications for theories of
binary evolution \cite{eb01b}.  In this paper we report in detail on
the observing system, analysis procedures, sensitivity and
completeness. We discuss detections of previously known pulsars and
present the new sample of slow pulsars, including timing results for
those with  solutions.

\section{Observations and Analysis}
\subsection{Hardware Configuration and Survey Observations}
The 64-m Parkes radio telescope was used with the 13-beam 21-cm
receiver \cite{swb+96} which provides 300 MHz of bandwidth and a
system temperature of $\sim$21 K. Signals from the two orthogonal
polarisations of each beam were mixed with a local oscillator before
being fed to an array of 26 96-channel filterbanks. Each filterbank
channel was 3 MHz wide and the band was centred at a frequency of
1374~MHz. The detected signals from corresponding polarisation pairs
in each channel were summed and high pass filtered (with a time
constant of $\sim$0.9~s; \pcite{mlc+00}) before being integrated and
one-bit sampled every 125 $\mu$s. The data stream was written to
magnetic tape (DLT 7000) for offline processing, as well as being made
available to online interference monitoring software in near-real-time
via the computer network. With the exception of the sampling interval,
the system was identical to that used for the Galactic plane survey
\cite{mlc+01}.

The receiver feeds are arranged in such a way as to allow coverage of
the sky in a hexagonal grid, with beams overlapping at their approximate
half-power points (7\arcmin\ from the beam centre). A group of four
pointings results in the uniform coverage of a roughly circular shape
$\sim1$ degree in radius which in turn can be efficiently tessellated
(see Figure \ref{fig:mbshape}).  The region enclosed by $5\degr < |b|
< 15\degr$ and $-100\degr < l < 50\degr$ was covered in 4,764 265-s
proposed pointings, amounting to only 14.6 days of integration time.
Most of these pointings were observed in several week-long observing
runs between August 1998 and August 1999.

\subsection{Search Analysis Procedure}
The processing of the 64-tape $\sim1.6$~terabyte data set was performed on
the Swinburne Supercluster, a network of 64 Compaq Alpha
workstations. Before searching for pulsars, each beam was analysed for
the presence of powerful signals that appeared in only a few
filterbank channels, a common type of interference signal. When such
signals were present, samples in the culprit channels were zeroed, a
process that does not incur too much loss of sensitivity since this
varies as the square root of the effective bandwidth.  In addition,
broad-band periodic signals that appear in large numbers of beams in
any given 30-minute period were detected and logged to a file for
later reference.  

To correct for the effects of non-linearity in the dispersion
relation, data from the 96 filterbank channels were padded with 32
dummy channels in such a way as to allow linear de-dispersion of the
resulting 128 channels, as used by the Galactic plane survey collaboration.
This enabled the use of the fast `tree'
algorithm of \scite{tay74} to partially de-disperse the data into eight
trial dispersion measures in each of sixteen sub-bands.  Whilst the
linearity of dispersion with respect to channel number would enable
full de-dispersion (that is, 128 trial DMs with no sub-band divisions),
in order to limit storage requirements and to allow the recording of
frequency-resolved pulse profiles to aid in suspect scrutiny,
application of the algorithm was stopped after the production of eight
DMs. 

The tree algorithm produces trial DMs up to the `diagonal' DM of
$17.0$~cm$^{-3}$~pc, where the dispersion delay across one sub-band in
units of samples is equal to the number of channels used to form
it. It should be noted that in previous surveys where the linear
dispersion approximation was acceptable for the tree stage, this
parameter was approximately equal to the DM at which the smearing
induced in each channel was one sample interval. The latter parameter
is commonly quoted in conjunction with the sampling interval to give
an indication of the time resolution available to pulsars of various
DMs in a pulsar survey. For the present survey this value varies from
$9.4$ to $17.5$ cm$^{-3}$~pc depending on the centre frequency of the
channel, and for evaluation purposes one should use the geometric mean
of $13.4$ cm$^{-3}$~pc. The tree algorithm was extended to also
produce time series for 1--2 times the diagonal DM, and beyond this
value the sample interval was doubled by summing of samples before
re-application of the algorithm, and the process repeated to produce
time series with 2--4, 4--8, 8--16 and 16--32 times the diagonal DM.

The periodicity search itself was based on that of the Parkes Southern
Pulsar Survey \cite{mld+96}, generalized and modified to handle the
large number of spurious interference signals present in the multibeam
data. Time series were constructed at 375 trial values of dispersion
measure from 0 to $562.5$ cm$^{-3}$~pc by summing partially
de-dispersed sub-bands in the nearest DM with the appropriate time
offsets. The trial DMs were spaced in such a way that the effective
smearing induced due to the difference between the DM of a pulsar and
the nearest trial DM was no more than twice that induced by the finite
width of individual filterbank channels.  The time series were
filtered with a boxcar of width 2048~ms to remove the effects of
receiver noise and gain variations during the course of the
observation, before being Fourier transformed and detected to form the
fluctuation power spectrum.

For signals with frequencies lying on the boundary between two
spectral bins the result is two components of equal magnitude and
opposite sign in the adjacent bins.  To maintain sensitivity to such
signals we also computed the difference of each bin and its neighbours
and used half the squared magnitude of the results as alternative
estimates of spectral power. For each bin the highest of the three
power values computed was chosen for use in the final power
spectrum. In the case of the zero-DM time series, this spectrum was
checked for the occurrence of any signal with a frequency close to one
earlier logged as a broad band interference signal contemporaneous
with this observation. Should such a signal be present, its exact
extent in the spectrum was assessed and the corresponding bins zeroed
in this and all other power spectra searched in this beam.  The
spectrum above a frequency of $1/12$~Hz was then searched for
significant spikes compared to a local mean (to compensate for the
overall redness of the spectrum). Harmonics were summed and the
process repeated for up to 16 harmonics to maintain sensitivity to
signals with short duty cycles. Significant signals at any level of
harmonic summing were recorded and after all trial dispersion measures
had been searched the set of signals was correlated into a number of
candidates, each covering signals of similar pulse period occurring
at multiple trial DMs. The top 99
candidates in each beam were subject to a fine search (by means of
maximisation of signal to noise ratio, S/N) in period and dispersion
measure around the best values found in the spectral search.
Pertinent information including the resulting best profile, grey scale
maps of pulse profiles as a function of time and radio frequency and
of signal to noise as a function of period and dispersion measure were
saved to disk.

\subsection{Suspect Scrutiny, Confirmation and Timing Observations}
The final stage of analysis was human viewing. The large number of
beams and the prevalence of interference signals presented
considerable complications to the viewing process due to the volume of
candidates produced. In previous surveys (e.g. \pcite{mld+96})
candidates of similar period occurring in multiple beams
contemporaneously were generally taken as interference signals and
ignored. In the case of results from this survey, the plethora of
interference signals across the spectrum resulted in the
misinterpretation of many pulsars as interference. It was found that
this limited the applicability of this approach to the handful of
periods that appeared more than $\sim250$ times on any tape. All
remaining candidates with signal to noise ratios greater than eight
(of which there were several hundred thousand) were then scrutinized
by a human viewer and promising signals scheduled for confirmation by
re-observation.

\begin{figure}
\centerline{\psfig{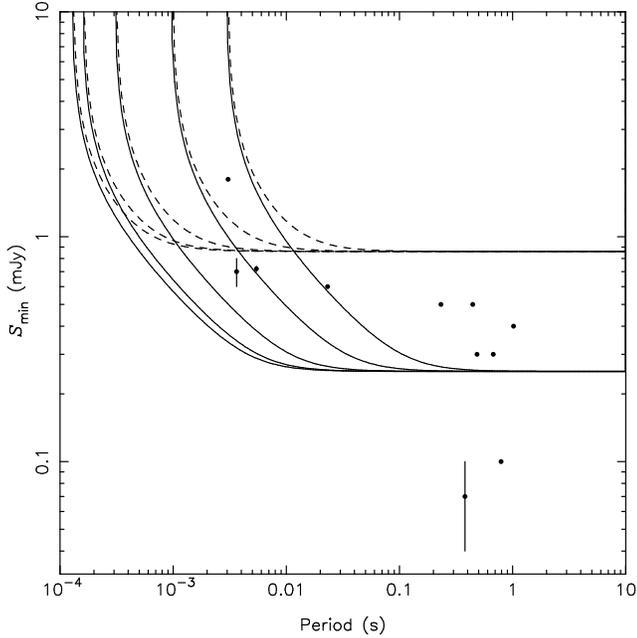}}
\caption{Estimated minimum detectable mean flux density ($S_{\rm
min}$) as a function of pulse period for intrinsic pulse widths of
$10\degr$ (solid lines) and $90\degr$ (dashed lines) at dispersion
measures of 0, 10, 30, 100 and 300 cm$^{-3}$~pc (in order of
increasing $S_{\rm min}$ for a given pulse period). Points represent
undetected pulsars which lie within $10\arcmin$ of an observed beam,
where flux density measurements have been published. Flux densities
published without uncertainties are plotted without error bars and in
such cases the relative uncertainty is probably around 50 per cent.}
\label{fig:sensitivity}
\end{figure}

\begin{figure}
\centerline{\psfig{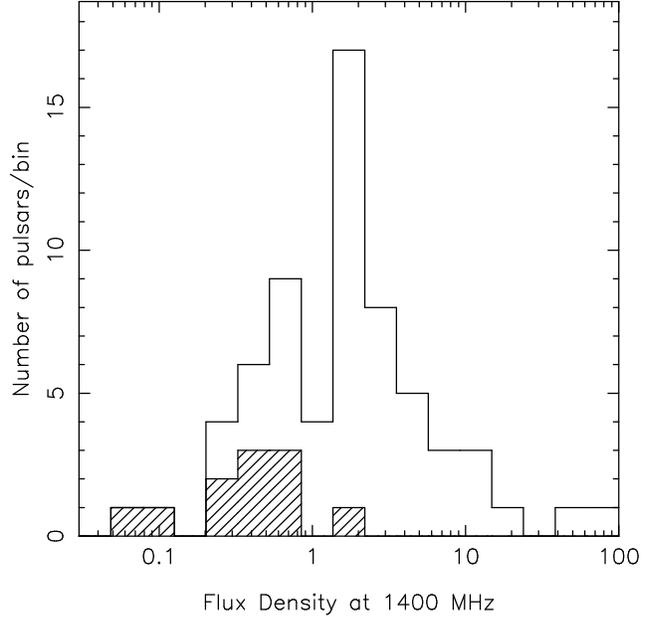}}
\caption{Histograms depicting distribution in flux density of all
previously known pulsars with published flux density with processed
beams centred less than 10$\arcmin$ away. The subset of such pulsars
that were undetected in the survey is represented by the hatched
regions.}
\label{fig:fluxhist}
\end{figure}

\begin{table*}
\begin{minipage}{11cm}
\caption{Detected previously known pulsars}
\label{tab:knowndet}
\begin{tabular}{llllllll}
Name & $P$ & DM & $S_{1400}$ & $l$ & $b$ & $\Delta$ pos & S/N \\ 
 & (s) & (cm$^{-3}$ pc) & (mJy) & (\degr) & (\degr) & (\arcmin) & \\
\hline
\\
B0743--53 & 0.215 & 121.5 &\dotfill & $-93.3$ & $-14.3$ & 7.3 & 85.7 \\ 
B0808--47 & 0.547 & 228.3 &3.00 & $-96.7$ & $-8.0$ & \dotfill & 46.4 \\ 
B0839--53 & 0.721 & 156.5 &2.00 & $-89.2$ & $-7.1$ & 5.2 & 55.1 \\ 
B0855--61 & 0.963 & 95 &\dotfill & $-81.4$ & $-10.4$ & 4.2 & 20.6 \\ 
B0901--63 & 0.660 & 76 &\dotfill & $-79.6$ & $-11.1$ & 8.7 & 25.2 \\ 
\\
B0950--38 & 1.374 & 167 &\dotfill & $-91.3$ & $12.0$ & 8.2 & 12.9 \\ 
B0957--47 & 0.670 & 92.3 &\dotfill & $-84.3$ & $5.4$ & 5.2 & 34.7 \\ 
B1001--47 & 0.307 & 98.5 &\dotfill & $-84.0$ & $6.1$ & 2.9 & 31.4 \\ 
J1036--4926 & 0.510 & 136.5 &\dotfill & $-78.5$ & $7.7$ & 5.8 & 10.8 \\ 
J1045--4509 & 0.007 & 58.1 &3.00 & $-79.1$ & $12.3$ & 0.0 & 37.6 \\ 
\\
J1047--6709 & 0.198 & 116.2 &4.00 & $-68.7$ & $-7.1$ & 3.6 & 87.0 \\ 
B1055--52 & 0.197 & 30.1 &\dotfill & $-74.0$ & $6.6$ & \dotfill & 51.0 \\ 
B1110--65 & 0.334 & 249.1 &\dotfill & $-66.8$ & $-5.2$ & 5.2 & 38.1 \\ 
B1110--69 & 0.820 & 148.4 &\dotfill & $-65.6$ & $-8.2$ & 5.0 & 18.1 \\ 
B1119--54 & 0.536 & 205.1 &\dotfill & $-69.9$ & $5.9$ & 4.2 & 44.7 \\ 
\\
J1123--4844 & 0.245 & 92.9 &\dotfill & $-71.7$ & $11.6$ & 6.7 & 36.1 \\ 
J1126--6942 & 0.579 & 55.3 &\dotfill & $-64.4$ & $-8.0$ & 3.3 & 15.9 \\ 
B1133--55 & 0.365 & 85.2 &4.00 & $-67.7$ & $5.9$ & 5.6 & 100.7 \\ 
J1210--5559 & 0.280 & 174.3 &2.10 & $-62.9$ & $6.4$ & 7.4 & 33.1 \\ 
B1232--55 & 0.638 & 100 &1.00 & $-59.4$ & $7.5$ & 19.2 & 22.7 \\ 
\\
B1236--68 & 1.302 & 94.1 &\dotfill & $-58.1$ & $-5.7$ & 11.1 & 32.7 \\ 
B1309--53 & 0.728 & 133 &\dotfill & $-54.0$ & $8.7$ & 2.2 & 20.3 \\ 
B1309--55 & 0.849 & 135.1 &\dotfill & $-54.0$ & $7.5$ & 0.8 & 102.4 \\ 
B1317--53 & 0.280 & 97.6 &\dotfill & $-52.7$ & $8.6$ & 4.8 & 43.2 \\ 
B1325--49 & 1.479 & 118 &\dotfill & $-50.9$ & $13.1$ & 6.1 & 17.1 \\ 
\\
J1350--5115 & 0.296 & 90.4 &\dotfill & $-47.8$ & $10.5$ & 4.9 & 25.1 \\ 
B1352--51 & 0.644 & 112.1 &\dotfill & $-47.0$ & $9.7$ & 3.2 & 40.9 \\ 
B1359--51 & 1.380 & 39 &\dotfill & $-45.9$ & $9.9$ & 63.7 & 71.0 \\ 
J1403--7646$^\ddag$ & 1.306 & 100.6 &\dotfill & $-52.9$ & $-14.5$ & 2.2 & 14.5 \\ 
B1417--54 & 0.936 & 129.6 &\dotfill & $-44.2$ & $6.4$ & 6.1 & 30.7 \\ 
\\
B1426--66 & 0.785 & 65.3 &6.00 & $-47.3$ & $-5.4$ & 6.7 & 126.2 \\ 
B1451--68 & 0.263 & 8.7 &80.00 & $-46.1$ & $-8.5$ & 7.0 & 242.9 \\ 
B1454--51 & 1.748 & 35.1 &\dotfill & $-37.9$ & $6.7$ & 4.4 & 27.1 \\ 
B1503--66 & 0.356 & 129.8 &\dotfill & $-44.1$ & $-7.3$ & 1.9 & 66.5 \\ 
B1504--43 & 0.287 & 48.7 &\dotfill & $-32.7$ & $12.5$ & 4.4 & 107.5 \\ 
\\
B1507--44 & 0.944 & 84 &\dotfill & $-32.4$ & $11.7$ & \dotfill & 53.1 \\ 
B1510--48 & 0.455 & 49.3 &\dotfill & $-34.1$ & $7.8$ & 5.9 & 16.3 \\ 
B1524--39 & 2.418 & 46.8 &\dotfill & $-27.0$ & $14.0$ & 6.7 & 37.5 \\ 
B1556--44 & 0.257 & 56.3 &40.00 & $-25.5$ & $6.4$ & 7.0 & 71.3 \\ 
J1557--4258 & 0.329 & 144.5 &5.10 & $-24.7$ & $8.0$ & 5.6 & 59.5 \\ 
\\
J1603--7202 & 0.015 & 38.1 &2.9(2) & $-43.4$ & $-14.5$ & 3.6 & 16.1 \\ 
J1614--3937 & 0.407 & 152.4 &\dotfill & $-20.0$ & $8.2$ & 6.1 & 11.7 \\ 
B1620--42$^\dag$ & 0.365 & 295 &2.20 & $-21.1$ & $4.6$ & 13.8 & 12.5 \\ 
J1625--4048 & 2.355 & 145 &\dotfill & $-19.4$ & $5.9$ & 8.1 & 16.1 \\ 
B1630--59 & 0.529 & 134.9 &\dotfill & $-32.3$ & $-8.3$ & 1.5 & 50.9 \\ 
\\
B1641--68 & 1.786 & 43 &2.00 & $-38.2$ & $-14.8$ & 4.6 & 110.7 \\ 
B1647--52 & 0.635 & 179.1 &2.00 & $-25.0$ & $-5.2$ & 4.0 & 117.2 \\ 
B1647--528 & 0.891 & 164 &\dotfill & $-25.4$ & $-5.5$ & 5.8 & 64.1 \\ 
J1648--3256 & 0.719 & 128.3 &1.00 & $-10.4$ & $7.7$ & 5.5 & 21.0 \\ 
B1649--23 & 1.704 & 68.3 &1.10 & $-2.7$ & $12.5$ & \dotfill & 38.0 \\ 
\multicolumn{6}{l}{$^\dag$ --- Pulsar lies outside nominal survey region}\\
\multicolumn{6}{l}{$^\ddag$ --- Pulsar originally published with incorrect period; corrected period listed}
\end{tabular}
\end{minipage}\end{table*}
\begin{table*}
\begin{minipage}{11cm}
\contcaption{}
\begin{tabular}{llllllll}
Name & $P$ & DM & $S_{1400}$ & $l$ & $b$ & $\Delta$ pos & S/N \\ 
 & (s) & (cm$^{-3}$ pc) & (mJy) & (\degr) & (\degr) & (\arcmin) & \\
\hline
\\
B1700--18 & 0.804 & 48.3 &0.7(2) & $3.2$ & $13.6$ & 6.4 & 18.2 \\ 
J1700--3312 & 1.358 & 166.8 &\dotfill & $-8.9$ & $5.5$ & 9.1 & 21.5 \\ 
B1702--19 & 0.299 & 22.9 &8(3) & $3.2$ & $13.0$ & 7.1 & 56.6 \\ 
B1706--16 & 0.653 & 24.9 &4(2) & $5.8$ & $13.7$ & 6.5 & 231.4 \\ 
B1707--53 & 0.899 & 106.1 &\dotfill & $-24.3$ & $-8.5$ & 0.4 & 16.9 \\ 
\\
B1709--15 & 0.869 & 58.0 &0.7(2) & $7.4$ & $14.0$ & 4.4 & 15.9 \\ 
B1717--16 & 1.566 & 44.9 &1.1(4) & $7.4$ & $11.5$ & 3.2 & 19.5 \\ 
B1718--19 & 1.004 & 75.9 &0.30 & $4.9$ & $9.7$ & 5.2 & 9.1 \\ 
B1727--47 & 0.830 & 123.3 &12.00 & $-17.4$ & $-7.7$ & 2.3 & 582.4 \\ 
B1730--22 & 0.872 & 41.2 &2.2(3) & $4.0$ & $5.7$ & 5.6 & 64.5 \\ 
\\
J1730--2304 & 0.008 & 9.6 &3.0(4) & $3.1$ & $6.0$ & \dotfill & 48.7 \\ 
B1732--07 & 0.419 & 73.5 &1.7(2) & $17.3$ & $13.3$ & 3.5 & 78.4 \\ 
B1738--08 & 2.043 & 74.9 &1.4(4) & $17.0$ & $11.3$ & 4.9 & 91.3 \\ 
B1740--13 & 0.405 & 115 &0.50(10) & $12.7$ & $8.2$ & 3.9 & 24.0 \\ 
J1744--1134 & 0.004 & 3.1 &2.0(2) & $14.8$ & $9.2$ & 3.2 & 24.8 \\ 
\\
B1745--12 & 0.394 & 100.0 &2.0(3) & $14.0$ & $7.7$ & 3.8 & 106.1 \\ 
B1747--46 & 0.742 & 20.3 &10.00 & $-15.0$ & $-10.2$ & 6.5 & 159.0 \\ 
B1758--03 & 0.921 & 117.6 &0.70(10) & $23.6$ & $9.3$ & 4.1 & 30.3 \\ 
B1802+03 & 0.219 & 79.4 &\dotfill & $30.4$ & $11.7$ & 6.4 & 27.7 \\ 
B1804--08 & 0.164 & 112.8 &16(4) & $20.1$ & $5.6$ & 5.2 & 260.9 \\ 
\\
J1808+00 & 0.425 & 141 &\dotfill & $28.5$ & $9.8$ & 8.1 & 21.4 \\ 
J1808--0813 & 0.876 & 151.3 &2.00 & $20.6$ & $5.8$ & 7.1 & 34.9 \\ 
B1810+02 & 0.794 & 101.6 &0.30 & $30.7$ & $9.7$ & 4.7 & 14.6 \\ 
J1811+0702 & 0.462 & 54 &\dotfill & $34.7$ & $12.1$ & 2.2 & 21.0 \\ 
B1813--26$^\dag$ & 0.593 & 128.1 &\dotfill & $5.2$ & $-4.9$ & 7.8 & 17.9 \\ 
\\
B1813--36 & 0.387 & 94.4 &2.00 & $-3.2$ & $-9.4$ & 5.4 & 52.4 \\ 
J1817--3837$^\ddag$ & 0.384 & 102.8 &\dotfill & $-5.3$ & $-10.4$ & 7.4 & 57.6 \\ 
B1818--04$^\dag$ & 0.598 & 84.4 &8.0(6) & $25.5$ & $4.7$ & 12.0 & 104.6 \\ 
B1820--31 & 0.284 & 50.3 &2.5(6) & $2.1$ & $-8.3$ & 8.3 & 72.1 \\ 
B1821+05 & 0.753 & 67.2 &1.7(4) & $35.0$ & $8.9$ & 3.3 & 111.9 \\ 
\\
B1822+00 & 0.779 & 54.4 &0.40(10) & $30.0$ & $5.9$ & 1.4 & 26.6 \\ 
J1822--4209 & 0.457 & 72.5 &1.50 & $-8.1$ & $-12.8$ & 6.3 & 16.2 \\ 
B1839+09 & 0.381 & 49.1 &1.70(10) & $40.1$ & $6.3$ & 3.5 & 99.9 \\ 
B1842+14 & 0.375 & 41.2 &1.5(3) & $45.6$ & $8.1$ & 5.2 & 30.5 \\ 
B1845--19 & 4.308 & 18.3 &\dotfill & $14.8$ & $-8.3$ & 9.3 & 84.6 \\ 
\\
B1848+12 & 1.205 & 71 &0.50(10) & $44.5$ & $5.9$ & 5.5 & 21.3 \\ 
B1848+13 & 0.346 & 59.0 &1.4(3) & $45.0$ & $6.3$ & 8.1 & 32.1 \\ 
J1848--1414 & 0.298 & 134.4 &\dotfill & $19.9$ & $-5.8$ & 4.3 & 11.2 \\ 
B1851--14 & 1.147 & 130.1 &\dotfill & $20.5$ & $-7.2$ & 9.0 & 24.0 \\ 
J1852--2610 & 0.336 & 56.8 &1.40 & $9.5$ & $-11.9$ & 0.8 & 36.2 \\ 
\\
B1857--26 & 0.612 & 38.1 &13.0(10) & $10.3$ & $-13.5$ & 6.8 & 201.6 \\ 
B1900--06 & 0.432 & 195.7 &\dotfill & $28.5$ & $-5.7$ & 10.0 & 13.7 \\ 
J1901--0906$^\ddag$ & 1.782 & 72.7 &3.10 & $26.0$ & $-6.4$ & 4.7 & 98.6 \\ 
J1904--1224 & 0.751 & 118.2 &\dotfill & $23.3$ & $-8.5$ & 2.5 & 14.9 \\ 
B1907--03 & 0.505 & 205.7 &0.80 & $32.3$ & $-5.7$ & 3.6 & 22.1 \\ 
\\
B1911--04 & 0.826 & 89.4 &4.4(5) & $31.3$ & $-7.1$ & 4.8 & 233.4 \\ 
B1917+00 & 1.272 & 90.7 &0.8(2) & $36.5$ & $-6.2$ & 4.6 & 36.7 \\ 
B1923+04 & 1.074 & 101.8 &\dotfill & $41.0$ & $-5.7$ & 9.0 & 15.5 \\ 
J1929+00 & 1.167 & 33 &\dotfill & $37.7$ & $-8.3$ & 2.3 & 14.0 \\ 
J1938+0652 & 1.122 & 70 &\dotfill & $44.4$ & $-7.1$ & 3.8 & 29.2 \\ 
\\
B1942--00 & 1.046 & 58.1 &0.80(10) & $38.6$ & $-12.3$ & 6.2 & 21.5 \\ 
\multicolumn{6}{l}{$^\dag$ --- Pulsar lies outside nominal survey region}\\
\multicolumn{6}{l}{$^\ddag$ --- Pulsar originally published with incorrect period; corrected period listed}
\end{tabular}
\end{minipage}\end{table*}
\begin{table*}
\begin{minipage}{10cm}
\caption{Undetected previously known pulsars}
\begin{tabular}{lllllll}
Name & $P$ & DM & $S_{1400}$ &  $l$ & $b$ & $\Delta$ pos \\ 
 & (s) & (cm$^{-3}$ pc) & (mJy)  & (\degr) & (\degr)& (\arcmin)  \\
\hline
\\
B0923--58 & 0.740 & 57.7 &\dotfill & $-81.6$ & $-5.6$ & 15.7 \\ 
J1006--6311 & 0.836 & 196.0 &\dotfill & $-74.4$ & $-6.0$ & 21.6 \\ 
J1123--6651 & 0.233 & 111.2 &0.50 & $-65.5$ & $-5.4$ & 5.4 \\ 
J1130--6807 & 0.256 & 148.7 &\dotfill & $-64.5$ & $-6.4$ & 5.8 \\ 
J1137--6700 & 0.556 & 228.0 &1.10 & $-64.2$ & $-5.2$ & 17.7 \\ 
\\
J1143--5158 & 0.676 & 159.0 &0.30 & $-67.6$ & $9.5$ & 3.4 \\ 
J1225--5556 & 1.018 & 125.8 &0.40 & $-60.7$ & $6.7$ & 6.6 \\ 
J1356--5521 & 0.507 & 174.2 &1.50 & $-47.8$ & $6.3$ & 11.8 \\ 
B1503--51 & 0.841 & 61.0 &\dotfill & $-36.9$ & $5.5$ & 8.4 \\ 
J1604--7203 & 0.341 & 54.4 &\dotfill & $-43.3$ & $-14.6$ & 4.1 \\ 
\\
J1654--2713 & 0.792 & 92.4 &0.10 & $-5.0$ & $10.3$ & 3.1 \\ 
B1659--60 & 0.306 & 54 &\dotfill & $-30.2$ & $-11.4$ & 6.3 \\ 
B1700--32 & 1.212 & 109.6 &6.00 & $-8.2$ & $5.4$ & 21.0 \\ 
J1701--3006 & 0.005 & 115.6 &\dotfill & $-6.4$ & $7.3$ & 6.6 \\ 
J1732--1930 & 0.484 & 73.0 &0.30 & $6.4$ & $7.6$ & 5.3 \\ 
\\
B1740--03 & 0.445 & 30.2 &0.50 & $21.6$ & $13.4$ & 5.0 \\ 
J1740--5340 & 0.004 & 71.9 &\dotfill & $-21.8$ & $-12.0$ & 3.7 \\ 
B1745--56 & 1.332 & 58 &\dotfill & $-23.4$ & $-14.3$ & 50.1 \\ 
B1802--07 & 0.023 & 186.4 &0.60 & $20.8$ & $6.8$ & 7.6 \\ 
J1807+07 & 0.464 & 89 &\dotfill & $35.1$ & $13.3$ & 4.1 \\ 
\\
J1809--3547 & 0.860 & 193.8 &\dotfill & $-3.5$ & $-7.8$ & 7.6 \\ 
B1820--30A & 0.005 & 86.8 &0.72(2) & $2.8$ & $-7.9$ & 6.6 \\ 
B1820--30B & 0.379 & 87.0 &0.07(3) & $2.8$ & $-7.9$ & 6.6 \\ 
J1821+17 & 1.366 & 79 &\dotfill & $45.3$ & $14.2$ & 8.7 \\ 
B1821--24 & 0.003 & 119.8 &1.80 & $7.8$ & $-5.6$ & 6.4 \\ 
\\
J1822+0705 & 1.363 & 50 &\dotfill & $36.0$ & $9.7$ & 4.2 \\ 
J1822+11 & 1.787 & 112 &\dotfill & $39.9$ & $11.6$ & 7.4 \\ 
J1823--0154 & 0.760 & 135.9 &0.80 & $28.1$ & $5.3$ & 28.7 \\ 
J1834+10 & 1.173 & 62 &\dotfill & $40.6$ & $8.6$ & 8.5 \\ 
J1838+06 & 1.122 & 70 &\dotfill & $37.5$ & $6.1$ & 4.0 \\ 
\\
J1838+16 & 1.902 & 36 &\dotfill & $46.7$ & $10.3$ & 10.2 \\ 
J1859+1526 & 0.934 & 97.4 &\dotfill & $47.6$ & $5.2$ & 76.0 \\ 
J1911--1114 & 0.004 & 31.0 &0.70(10) & $25.1$ & $-9.6$ & 6.2 \\ 
J1933+07 & 0.437 & 170 &\dotfill & $44.8$ & $-5.6$ & 10.2 \\ 
J1941+1026 & 0.905 & 138.9 &\dotfill & $48.0$ & $-6.2$ & 15.9 \\ 
\\
J1947+10 & 1.111 & 149 &\dotfill & $49.0$ & $-7.3$ & 56.5 \\ 
J1950+05 & 0.456 & 71 &\dotfill & $44.9$ & $-10.6$ & 5.5 \\ 
\end{tabular}
\label{tab:knownnondet}
\end{minipage}\end{table*}

Human viewing of all suspects with S/N $>$ 8 was expected to be
incomplete in its selection due to the viewing speed necessitated by
the large number of candidates to be assessed. This method was used as
a first pass over the data, however after all data were processed and
the 30 GB result set assembled, a more complete candidate analysis
scheme was employed. A custom-written graphical software package
allowed for visual (and numerical) identification of the distribution
of candidates in a variety of parameters. Sets of candidates could be
trimmed by the graphical or command-driven selection and deletion of
interference signals, and the remaining candidate list subjected to
human scrutiny. A set of `macros' were developed for the deletion of
dozens of commonly appearing highly coherent interference periods as
well as all signals with large relative errors in dispersion measure
(a characteristic of terrestrial, non-dispersed interference) and any
candidate with a S/N less than 9 (or 9.5 for $P < 20$~ms). These
produced an order of magnitude reduction in the number of suspects to
be viewed, and a corresponding improvement in the accuracy and
completeness of scrutiny. Interference mitigation procedures employed
for each tape were also recorded as macros to allow for repeatability
and quantification of any selection effects imposed.

Those candidates confirmed by detection in a re-observation were added
to an ongoing program of pulsar timing of new
discoveries. Observations of typically 250~s were made with the centre
beam of the system described above, or more recently the
$2\times512\times0.5$-MHz filterbank to provide improved time resolution for
short period pulsars. For most pulsars at least one timing observation
was obtained at a frequency of 660~MHz to enable accurate measurement
of the dispersion measure.  In offline processing samples were
de-dispersed and folded at the predicted topocentric pulse period. The
resulting pulse profiles were fitted to a `standard' profile usually
produced by adding several prior observations of high signal to noise
ratio. The resulting phase offsets were used to produce barycentric
times-of-arrival (TOAs) which were then used in conjunction with the
TEMPO software package\footnote{http://pulsar.princeton.edu/tempo} to
fit for the relevant spin, astrometric and binary parameters
(see e.g. \pcite{tw89}).

\begin{figure*}
\centerline{\psfig{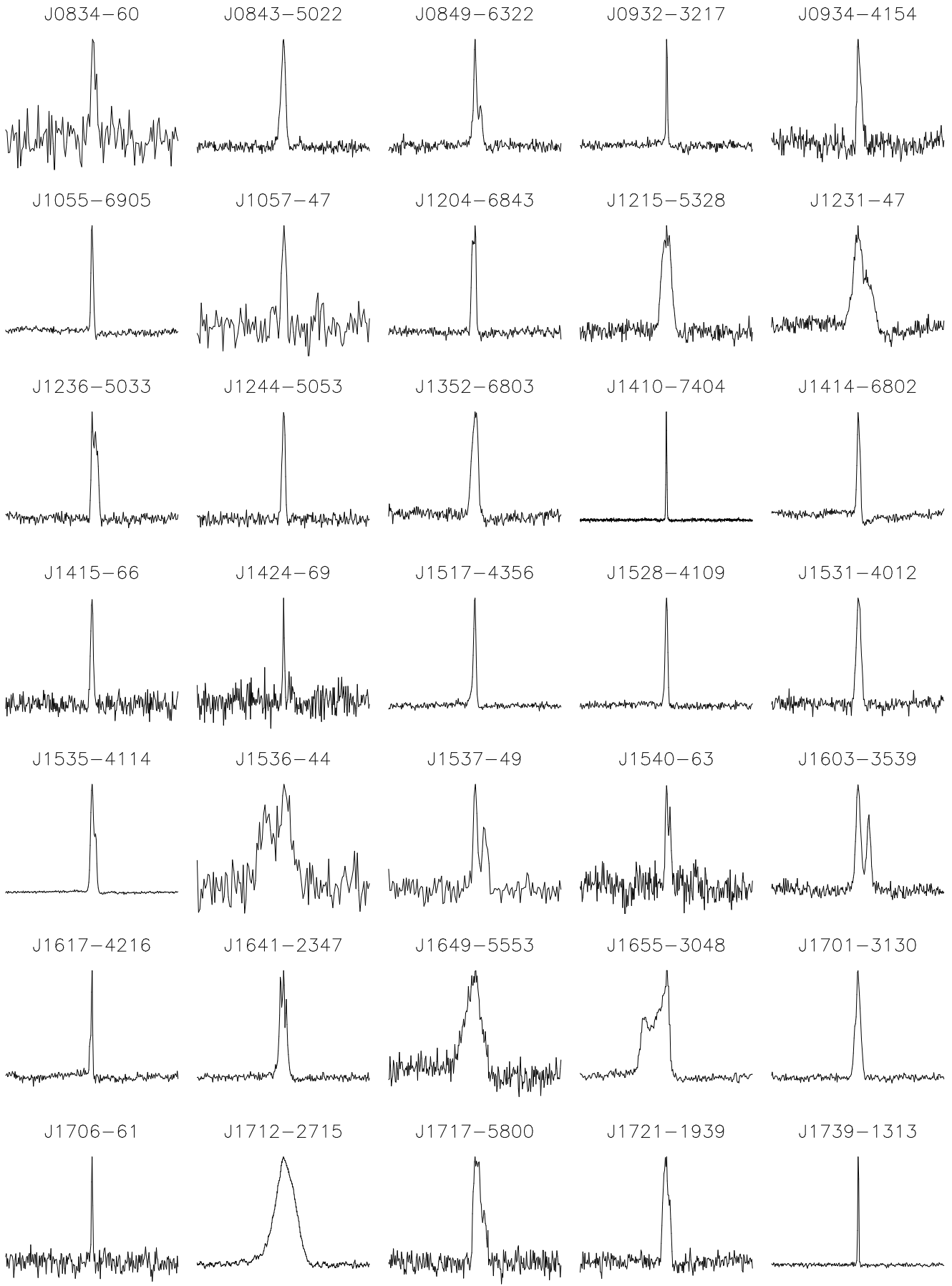}}
\caption{Pulse profiles for new slow pulsars}
\label{fig:profs}
\end{figure*}

\begin{figure*}
\centerline{\psfig{figure=profs1.ps,width=17.5cm}}
\contcaption{}
\end{figure*}

\section{Results and Discussion}
\subsection{Sky Coverage, Sensitivity and Re-Detections}

The survey was deemed complete in August 1999 after the observation
and successful processing of 60,852 beams in 4,702 pointings. In the
first observing run an error in the telescope control system resulted
in spurious rotations of the receiver feed, making the sky position 
corresponding to each recorded beam (except the centre beam) indeterminate
and reducing sensitivity by moving off (or on!) source midway through
observations. For this reason most of the region of sky surveyed under
these conditions were re-observed. 

Figure \ref{fig:coverage} shows the sky coverage achieved by the
survey. One ellipse is plotted for each group of four inter-meshing
pointings. Ellipses are shaded according to the number of observed and
processed pointings in the group. Most groups are either medium grey
(for standard once-only coverage), black (for those areas observed
twice due to position uncertainty as discussed above) or white (for
unobserved groups). Other shades reflect varying numbers of observed
and processed beams in the group. A total of 4,465 of the proposed
pointings were observed at least once, yielding a metric of completeness
of 4,465$/$4,764~$=$~94 per cent.

\begin{figure*}
\centerline{\psfig{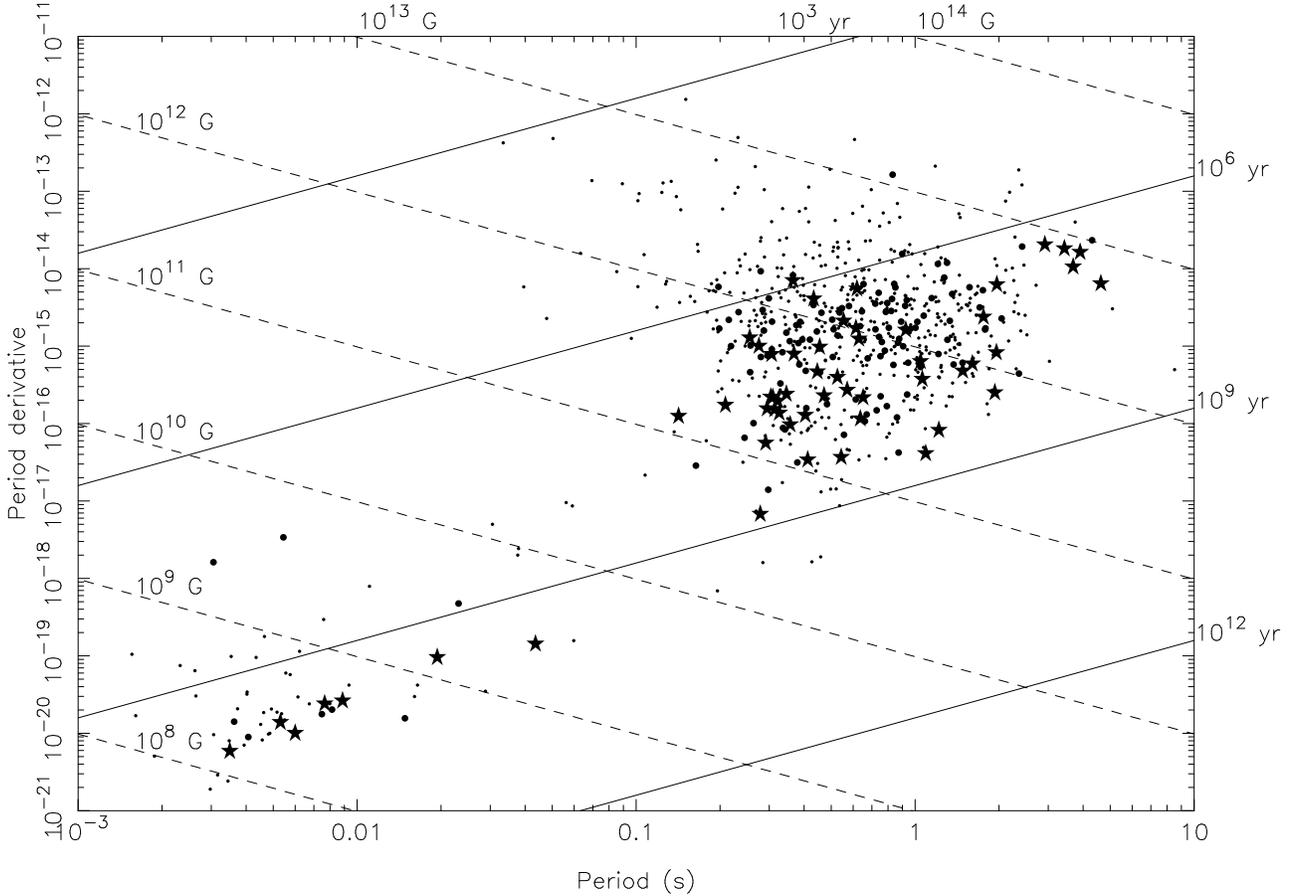}}
\caption{Distribution in pulse period and period derivative of
new pulsars (stars),previously known pulsars in the survey
region (large dots) and all other known pulsars (small dots),
where such values have been measured.
Also plotted are solid and dashed lines of constant characteristic
age ($\tau_c \equiv P/2\dot{P}$) and inferred surface magnetic
field strength ($B = 3.2\times10^{19} $ G s$^{-1/2}\sqrt{P\dot{P}}$) respectively.}
\label{fig:ppdot}
\end{figure*}

The sensitivity of a pulsar survey is derived from the radiometer
equation, altered to take into account the detection of periodic
signals. It can be expressed as follows,
\begin{equation}
S_{\rm min} = \frac{\alpha\beta (T_{\rm rec} + T_{\rm sky})}
	           {G (N_{\rm p}\Delta\nu t_{\rm obs})^{1/2}}
	      \left(\frac{\delta}{1-\delta}\right)^{1/2}
\end{equation}
\noindent where $S_{\rm min}$ is the minimum detectable mean flux
density, $\alpha$ is the threshold S/N, $\beta$ is a dimensionless
correction factor for system losses, $T_{\rm rec}$ and $T_{\rm sky}$
are the receiver and sky noise temperatures, $G$ is the telescope
gain, $N_{\rm p}$ is the number of polarisations, $\Delta\nu$ is the
observing bandwidth, $t_{\rm obs}$ is the integration time, $W$ is the
effective pulse width in time units and $P$ is the pulse period
\cite{dtws85}. The effective pulse width ($W=\delta P$ where $\delta$ is
the observed duty cycle) is computed as the quadrature sum of the
intrinsic pulse width and pulse broadening terms due to such effects
as dispersion smearing, scatter-broadening and the finite sampling
interval.

The system characteristics of the multibeam receiver vary from beam to
beam and we use here averages of the values presented at the
instrument web
page\footnote{http://www.atnf.csiro.au/research/multibeam/lstavele/description.html},
yielding a receiver temperature of 21~K and a gain of
0.64~K~Jy$^{-1}$.  The dimensionless parameter $\beta$ embodies the
loss due to one-bit digitisation ($\sqrt{\pi/2}\simeq 1.25$) and other
system losses, which we treat collectively with $\beta=1.5$. Assuming
a typical sky temperature of 2~K, the calculated sensitivity as a
function of pulse period is plotted in Figure \ref{fig:sensitivity}
for several dispersion measures and pulse widths. Included in the
effective width calculation are the dispersion smearing in filterbank
channels and the sampling interval. It should be noted that the
sensitivity derived is that available at the centre of the beam.  The
average beam response is approximately Gaussian \cite{mlc+01} with a
half-power width of 14.3\arcmin, meaning that pulsars lying near the
beam overlap points will be detected with half the sensitivity
calculated. We calculate that the mean sensitivity within the
half-power radius is 73 per cent of that at the centre of the beam,
meaning that for statistical sensitivity estimates, the curves in
Figure \ref{fig:sensitivity} should be raised about a tenth of a
decade. The resulting basic mean sensitivity to slow pulsars is
0.3--1.1 mJy for intrinsic pulse widths of 10\degr--90\degr.

The signal-to-noise ratio used in the sensitivity equation is
calculated in the time domain. We note that some pulsar surveys
(e.g. \pcite{cnt96,mlc+01}) have based their sensitivity analysis on
the so-called `spectral' S/N, computed from the amplitudes of
harmonics in the power spectrum.  Our search code computes such a
value for all candidates, however this is used only for the selection
of signals to be subjected to a fine search in the time domain, and the
threshold value used (5.0) is sufficiently low that we expect no loss
of significant candidates at this stage.  The (time domain) S/N
threshold of 8.0 imposed in the selection of candidates for human
viewing is also irrelevant for the sensitivity analysis since it found
that nearly all candidates re-detected in subsequent observations had
S/Ns greater than 10.0, despite attempts to re-detect large numbers of
candidates with S/N between 8.0 and 10.0.  We therefore use
$\alpha=10.0$ in the above analysis and in the curves plotted in
Figure \ref{fig:sensitivity}.

We caution that this analysis should only be taken as approximate. The
variation of a pulsar's flux due to scintillation adds a considerable
degree of uncertainty to its detectability, and particularly for
millisecond pulsars lack of time resolution and prevalence of
interference signals cause the human viewer to effectively adopt a
higher threshold S/N. In particular, whilst this analysis suggests
that this and similar surveys could be significantly sensitive to
sub-millisecond pulsars, we would treat such a claim with
skepticism. A sub-millisecond pulsar with $P=0.8$~ms and a moderately
low dispersion measure of 25~cm$^{-3}$~pc would experience $0.25$~ms
of dispersion smearing, resulting in a pulse profile with a width of
at least $110\degr$. Combined with the fact that only 4 or perhaps 8
bins would be present in the pulse profile, the human viewer is forced
to judge the origin of the signal essentially entirely upon the
reported S/N, which itself becomes subject to significant uncertainty
when the number of bins is so few. We therefore expect that the
standard sensitivity analysis underestimates the minimum detectable
flux density by a factor of a few for pulsars with periods shorter
than a millisecond. Likewise, the minimum detectable flux density may
be underestimated for very slow ($P>5$~s) pulsars due to the rather
short time constant (0.9~s) employed in the digitiser system. The
$P^{-1/2}$ duty cycle dependence of slow pulsars (e.g. \pcite{ran90})
aids the situation somewhat, but nevertheless it is expected that a
few percent of very slow pulsars have pulse widths greater than 0.5~s,
and these would experience significant S/N loss.

\begin{table*}
\begin{minipage}{17cm}
\caption{Dispersion measure, astrometric and spin parameters for new slow pulsars}
\label{tab:newparams1}
\begin{tabular}{lllllll}
Name & $\alpha$ (J2000) & $\delta$ (J2000) & $P$ & $P$ Epoch & $\dot{P}$ & DM \\ 
 & & & (s) & (MJD) & ($10^{-15}$) & (cm$^{-3}$ pc)\\ 
 \hline
& & & & & &\\
J0834--60 & 08$^{\mathrm h}$34$^{\mathrm m}$50(40) & --60\degr35(5)\arcmin & $0.384645$(6) & 51401.1 & \ldots
 & $20$(6)\\
J0843--5022 & 08$^{\mathrm h}$43$^{\mathrm m}$09\fs884(8) & --50\degr22\arcmin43\farcs10(8) & $0.2089556931527$(14) & 51500.0 & $0.17238$(14) & $178.47$(9)\\
J0849--6322 & 08$^{\mathrm h}$49$^{\mathrm m}$42\fs59(2) & --63\degr22\arcmin35\farcs0(1) & $0.367953256307$(5) & 51500.0 & $0.7908$(5) & $91.29$(9)\\
J0932--3217 & 09$^{\mathrm h}$32$^{\mathrm m}$39\fs15(6) & --32\degr17\arcmin14\farcs2(8) & $1.93162674308$(19) & 51500.0 & $0.250$(15) & $102.1$(8)\\
J0934--4154 & 09$^{\mathrm h}$34$^{\mathrm m}$58\fs20(3) & --41\degr54\arcmin19\farcs5(3) & $0.570409236430$(14) & 51650.0 & $0.269$(3) & $113.79$(16)\\
& & & & & &\\
J1055--6905 & 10$^{\mathrm h}$55$^{\mathrm m}$44\fs71(9) & --69\degr05\arcmin11\farcs4(4) & $2.9193969868$(3) & 51500.0 & $20.336$(15) & $142.8$(4)\\
J1057--47 & 10$^{\mathrm h}$57$^{\mathrm m}$45(30) & --47\degr57(5)\arcmin & $0.62830$(3) & 50989.1 & \ldots
 & $60$(8)\\
J1204--6843 & 12$^{\mathrm h}$04$^{\mathrm m}$36\fs72(1) & --68\degr43\arcmin17\farcs19(8) & $0.3088608620097$(19) & 51500.0 & $0.21708$(19) & $133.93$(9)\\
J1215--5328 & 12$^{\mathrm h}$15$^{\mathrm m}$00\fs62(7) & --53\degr28\arcmin31\farcs6(7) & $0.63641413680$(5) & 51500.0 & $0.115$(4) & $163.0$(5)\\
J1231--47 & 12$^{\mathrm h}$31$^{\mathrm m}$40(30) & --47\degr46(5)\arcmin & $1.8732$(3) & 51200.8 & \ldots
 & $31$(30)\\
& & & & & &\\
J1236--5033 & 12$^{\mathrm h}$36$^{\mathrm m}$59\fs15(1) & --50\degr33\arcmin36\farcs3(1) & $0.294759771191$(4) & 51500.0 & $0.1556$(4) & $105.02$(11)\\
J1244--5053 & 12$^{\mathrm h}$44$^{\mathrm m}$11\fs48(1) & --50\degr53\arcmin20\farcs6(1) & $0.275207111323$(4) & 51500.0 & $0.9998$(4) & $109.95$(12)\\
J1352--6803 & 13$^{\mathrm h}$52$^{\mathrm m}$34\fs45(4) & --68\degr03\arcmin37\farcs1(4) & $0.628902546380$(16) & 51650.0 & $1.234$(3) & $214.6$(2)\\
J1410--7404 & 14$^{\mathrm h}$10$^{\mathrm m}$07\fs370(5) & --74\degr04\arcmin53\farcs32(2) & $0.2787294436271$(15) & 51460.0 & $0.00674$(9) & $54.24$(6)\\
J1414--6802 & 14$^{\mathrm h}$14$^{\mathrm m}$25\fs7(1) & --68\degr02\arcmin58(1)\arcsec & $4.6301880619$(4) & 51650.0 & $6.39$(7) & $153.5$(6)\\
& & & & & &\\
J1415--66 & 14$^{\mathrm h}$15$^{\mathrm m}$25(50) & --66\degr19(5)\arcmin & $0.392480$(10) & 51396.2 & \ldots
 & $261$(6)\\
J1424--69 & 14$^{\mathrm h}$24$^{\mathrm m}$15(60) & --69\degr56(5)\arcmin & $0.333415$(8) & 51309.7 & \ldots
 & $123$(4)\\
J1517--4356 & 15$^{\mathrm h}$17$^{\mathrm m}$27\fs34(1) & --43\degr56\arcmin17\farcs9(2) & $0.650836871901$(6) & 51500.0 & $0.2155$(6) & $87.78$(12)\\
J1528--4109 & 15$^{\mathrm h}$28$^{\mathrm m}$08\fs033(8) & --41\degr09\arcmin28\farcs8(2) & $0.526556139140$(4) & 51500.0 & $0.3955$(4) & $89.50$(10)\\
J1531--4012 & 15$^{\mathrm h}$31$^{\mathrm m}$08\fs05(1) & --40\degr12\arcmin30\farcs9(4) & $0.356849312855$(6) & 51500.0 & $0.0963$(6) & $106.65$(12)\\
& & & & & &\\
J1535--4114 & 15$^{\mathrm h}$35$^{\mathrm m}$17\fs07(1) & --41\degr14\arcmin03\farcs1(3) & $0.432866133845$(6) & 51500.0 & $4.0705$(6) & $66.28$(14)\\
J1536--44 & 15$^{\mathrm h}$36$^{\mathrm m}$15(30) & --44\degr16(5)\arcmin & $0.46842$(6) & 51063.2 & \ldots
 & $110$(30)\\
J1537--49 & 15$^{\mathrm h}$37$^{\mathrm m}$30(30) & --49\degr09(5)\arcmin & $0.301313$(6) & 51402.3 & \ldots
 & $65$(4)\\
J1540--63 & 15$^{\mathrm h}$40$^{\mathrm m}$20(40) & --63\degr24(5)\arcmin & $1.63080$(16) & 51307.7 & \ldots
 & $160$(20)\\
J1603--3539 & 16$^{\mathrm h}$03$^{\mathrm m}$53\fs697(5) & --35\degr39\arcmin57\farcs1(3) & $0.1419085889640$(9) & 51650.0 & $0.12425$(17) & $77.5$(4)\\
& & & & & &\\
J1617--4216 & 16$^{\mathrm h}$17$^{\mathrm m}$23\fs38(5) & --42\degr16\arcmin59(1)\arcsec & $3.42846630955$(13) & 51500.0 & $18.129$(15) & $163.6$(5)\\
J1641--2347 & 16$^{\mathrm h}$41$^{\mathrm m}$18\fs04(6) & --23\degr47\arcmin36(6)\arcsec & $1.091008429855$(16) & 51500.0 & $0.0411$(15) & $27.7$(3)\\
J1649--5553 & 16$^{\mathrm h}$49$^{\mathrm m}$31\fs1(1) & --55\degr53\arcmin40(2)\arcsec & $0.61357070436$(7) & 51650.0 & $1.698$(16) & $180.4$(11)\\
J1655--3048 & 16$^{\mathrm h}$55$^{\mathrm m}$24\fs53(2) & --30\degr48\arcmin42(1)\arcsec & $0.542935874228$(9) & 51500.0 & $0.0366$(9) & $154.3$(3)\\
J1701--3130 & 17$^{\mathrm h}$01$^{\mathrm m}$43\fs513(5) & --31\degr30\arcmin36\farcs7(4) & $0.2913414710251$(12) & 51500.0 & $0.05596$(12) & $130.73$(6)\\
\end{tabular}
\end{minipage}
\end{table*}
\begin{table*}
\begin{minipage}{17cm}
\contcaption{}
\begin{tabular}{lllllll}
Name & $\alpha$ (J2000) & $\delta$ (J2000) & $P$ & $P$ Epoch & $\dot{P}$ & DM \\ 
 & & & (s) & (MJD) & ($10^{-15}$) & (cm$^{-3}$ pc)\\ 
 \hline
& & & & & &\\
J1706--61 & 17$^{\mathrm h}$06$^{\mathrm m}$40(40) & --61\degr11(5)\arcmin & $0.361922$(8) & 51308.7 & \ldots
 & $78$(6)\\
J1712--2715 & 17$^{\mathrm h}$12$^{\mathrm m}$11\fs71(1) & --27\degr15\arcmin53(2)\arcsec & $0.255359660118$(3) & 51500.0 & $1.2793$(3) & $92.64$(13)\\
J1717--5800 & 17$^{\mathrm h}$17$^{\mathrm m}$35\fs65(2) & --58\degr00\arcmin05\farcs4(3) & $0.321793346869$(6) & 51650.0 & $0.1957$(10) & $125.22$(14)\\
J1721--1939 & 17$^{\mathrm h}$21$^{\mathrm m}$46\fs61(4) & --19\degr39\arcmin49(5)\arcsec & $0.404039751280$(15) & 51500.0 & $0.1283$(15) & $103$(2)\\
J1739--1313 & 17$^{\mathrm h}$39$^{\mathrm m}$57\fs821(6) & --13\degr13\arcmin18\farcs6(4) & $1.215697613611$(9) & 51500.0 & $0.0817$(9) & $58.2$(5)\\
& & & & & &\\
J1741--2019 & 17$^{\mathrm h}$41$^{\mathrm m}$06\fs87(3) & --20\degr19\arcmin24(5)\arcsec & $3.90450636119$(13) & 51500.0 & $16.260$(13) & $74.9$(4)\\
J1742--4616 & 17$^{\mathrm h}$42$^{\mathrm m}$26\fs10(2) & --46\degr16\arcmin53\farcs5(4) & $0.412401047219$(7) & 51650.0 & $0.0338$(12) & $115.96$(14)\\
J1743--4212 & 17$^{\mathrm h}$43$^{\mathrm m}$05\fs223(5) & --42\degr12\arcmin02\farcs4(2) & $0.3061669878595$(16) & 51650.0 & $0.7834$(4) & $131.94$(5)\\
J1744--1610 & 17$^{\mathrm h}$44$^{\mathrm m}$16\fs534(7) & --16\degr10\arcmin35\farcs8(8) & $1.757205868816$(16) & 51500.0 & $2.3767$(16) & $66.67$(14)\\
J1745--0129 & 17$^{\mathrm h}$45$^{\mathrm m}$02\fs06(1) & --01\degr29\arcmin18\farcs1(4) & $1.045406855598$(18) & 51650.0 & $0.631$(4) & $90.1$(11)\\
& & & & & &\\
J1802+0128 & 18$^{\mathrm h}$02$^{\mathrm m}$27\fs45(2) & +01\degr28\arcmin23\farcs7(4) & $0.554261603931$(10) & 51650.0 & $2.109$(3) & $97.97$(12)\\
J1805--0619 & 18$^{\mathrm h}$05$^{\mathrm m}$31\fs436(9) & --06\degr19\arcmin45\farcs4(4) & $0.454650713078$(7) & 51650.0 & $0.9690$(13) & $146.22$(9)\\
J1806+10 & 18$^{\mathrm h}$06$^{\mathrm m}$50(20) & +10\degr24(5)\arcmin & $0.484285$(15) & 51259.8 & \ldots
 & $58$(6)\\
J1808--3249 & 18$^{\mathrm h}$08$^{\mathrm m}$04\fs48(2) & --32\degr49\arcmin34(1)\arcsec & $0.364912241765$(10) & 51500.0 & $7.0494$(10) & $147.37$(19)\\
J1809--0743 & 18$^{\mathrm h}$09$^{\mathrm m}$35\fs92(1) & --07\degr43\arcmin01\farcs4(5) & $0.313885674748$(5) & 51650.0 & $0.1521$(9) & $240.70$(14)\\
& & & & & &\\
J1811--0154 & 18$^{\mathrm h}$11$^{\mathrm m}$19\fs88(3) & --01\degr54\arcmin30\farcs9(7) & $0.92494482303$(4) & 51650.0 & $1.608$(6) & $148.1$(3)\\
J1819+1305 & 18$^{\mathrm h}$19$^{\mathrm m}$56\fs22(4) & +13\degr05\arcmin14\farcs2(7) & $1.06036354400$(6) & 51650.0 & $0.373$(9) & $64.9$(4)\\
J1824--25 & 18$^{\mathrm h}$24$^{\mathrm m}$15(20) & --25\degr36(5)\arcmin & $0.223319$(3) & 51067.5 & \ldots
 & $155$(3)\\
J1832--28 & 18$^{\mathrm h}$32$^{\mathrm m}$30(20) & --28\degr43(5)\arcmin & $0.199300$(3) & 51064.3 & \ldots
 & $127$(3)\\
J1837+1221 & 18$^{\mathrm h}$37$^{\mathrm m}$07\fs12(4) & +12\degr21\arcmin54\farcs0(6) & $1.96353198352$(12) & 51650.0 & $6.200$(16) & $100.6$(4)\\
& & & & & &\\
J1837--1837 & 18$^{\mathrm h}$37$^{\mathrm m}$54\fs25(1) & --18\degr37\arcmin08(2)\arcsec & $0.618357697387$(16) & 51500.0 & $5.4950$(12) & $100.74$(13)\\
J1842+1332 & 18$^{\mathrm h}$42$^{\mathrm m}$29\fs96(6) & +13\degr32\arcmin01\farcs5(9) & $0.47160357893$(3) & 51650.0 & $0.229$(7) & $102.5$(7)\\
J1848+12 & 18$^{\mathrm h}$48$^{\mathrm m}$30(20) & +12\degr50(5)\arcmin & $0.75473$(7) & 51316.7 & \ldots
 & $139$(20)\\
J1855--0941 & 18$^{\mathrm h}$55$^{\mathrm m}$15\fs68(3) & --09\degr41\arcmin02(1)\arcsec & $0.34540115992$(4) & 51500.0 & $0.240$(3) & $152.2$(3)\\
J1857--1027 & 18$^{\mathrm h}$57$^{\mathrm m}$26\fs45(5) & --10\degr27\arcmin01(2)\arcsec & $3.6872190477$(3) & 51650.0 & $10.55$(6) & $108.9$(7)\\
& & & & & &\\
J1901--1740 & 19$^{\mathrm h}$01$^{\mathrm m}$18\fs03(6) & --17\degr40\arcmin00(6)\arcsec & $1.95685759005$(16) & 51500.0 & $0.823$(16) & $24.4$(6)\\
J1919+0134 & 19$^{\mathrm h}$19$^{\mathrm m}$43\fs62(3) & +01\degr34\arcmin56\farcs5(7) & $1.60398355528$(6) & 51650.0 & $0.589$(11) & $191.9$(4)\\
J1943+0609 & 19$^{\mathrm h}$43$^{\mathrm m}$29\fs132(5) & +06\degr09\arcmin57\farcs6(1) & $0.446226281658$(3) & 51650.0 & $0.4659$(6) & $70.76$(6)\\
J1947+0915 & 19$^{\mathrm h}$47$^{\mathrm m}$46\fs22(5) & +09\degr15\arcmin08\farcs0(8) & $1.48074382424$(9) & 51650.0 & $0.478$(16) & $94$(4)\\
J1956+0838 & 19$^{\mathrm h}$56$^{\mathrm m}$52\fs26(2) & +08\degr38\arcmin16\farcs8(4) & $0.303910924347$(7) & 51650.0 & $0.2199$(13) & $68.2$(13)\\
& & & & & &\\
J2007+0809 & 20$^{\mathrm h}$07$^{\mathrm m}$13\fs5(1) & +08\degr09\arcmin33(2)\arcsec & $0.32572436605$(5) & 51650.0 & $0.137$(7) & $53.9$(10)\\
\end{tabular}
\end{minipage}
\end{table*}
\begin{table*}
\begin{minipage}{15cm}
\caption{Detection parameters, pulse widths and derived parameters for new slow pulsars}
\label{tab:newparams2}
\begin{tabular}{llllllllllll}
Name & $\Delta$ pos & S/N& $w_{50}$ & $w_{10}$ & $l$ & $b$ & $d$ & $|z|$ & $\tau_c$ & $B$ & $\dot{E}$\\
 &   (\arcmin) & & (\degr) & (\degr) & (\degr) & (\degr) & (kpc) & (kpc)  & (Myr)& $(10^{12}$ G) & ($10^{30}$~erg~s$^{-1}$)\\
\hline
& & & & & & & & & & \\
J0834--60 & \ldots & 13.9 & \ldots &\ldots &$-83.9$ & $-11.9$ &$0.5$ &$0.10$ &\ldots &\ldots &\ldots 
\\
J0843--5022 & \ldots & 10.1 & 11.0 &50.0 &$-91.5$ & $-4.9$ &$7.7$ &$0.66$ &19.2 &0.19 &$746$
\\
J0849--6322 & 3.9 & 13.2 & 7.0 &166.3 &$-80.6$ & $-12.2$ &$>8.4$ &$>1.8$ &7.37 &0.55 &$627$
\\
J0932--3217 & 5.9 & 17.1 & 3.4 &7.6 &$-98.7$ & $14.1$ &$3.8$ &$0.93$ &122 &0.70 &$1.37$
\\
J0934--4154 & 4.4 & 10.4 & 10.1 &\ldots &$-91.6$ & $7.4$ &$3.2$ &$0.41$ &33.5 &0.40 &$57.3$
\\
& & & & & & & & & & \\
J1055--6905 & 3.5 & 17.8 & 5.8 &10.7 &$-67.1$ & $-8.5$ &$>12$ &$>1.8$ &2.27 &7.8 &$32.3$
\\
J1057--47 & \ldots & 16.3 & 11.5 &\ldots &$-76.0$ & $10.7$ &$3.0$ &$0.56$ &\ldots &\ldots &\ldots 
\\
J1204--6843 & 5.5 & 18.5 & 10.6 &15.2 &$-61.3$ & $-6.2$ &$5.7$ &$0.61$ &22.5 &0.26 &$291$
\\
J1215--5328 & 8.1 & 12.9 & 22.6 &\ldots &$-62.5$ & $9.0$ &$>11$ &$>1.8$ &87.4 &0.27 &$17.7$
\\
J1231--47 & \ldots & 28.5 & 30.7 &\ldots &$-60.5$ & $15.0$ &$1.6$ &$0.42$ &\ldots &\ldots &\ldots 
\\
& & & & & & & & & & \\
J1236--5033 & 3.2 & 13.2 & 14.7 &20.8 &$-59.4$ & $12.2$ &$>8.3$ &$>1.8$ &30.0 &0.22 &$240$
\\
J1244--5053 & 4.3 & 12.4 & 8.1 &\ldots &$-58.2$ & $12.0$ &$>8.5$ &$>1.8$ &4.36 &0.53 &$1894$
\\
J1352--6803 & 3.5 & 32.7 & 16.0 &188.9 &$-51.4$ & $-5.9$ &$14$ &$1.4$ &8.07 &0.89 &$196$
\\
J1410--7404 & 5.7 & 30.4 & 2.3 &4.5 &$-51.7$ & $-12.0$ &$2.1$ &$0.45$ &655 &0.044 &$12.3$
\\
J1414--6802 & 2.0 & 27.3 & 7.8 &14.0 &$-49.4$ & $-6.4$ &$6.6$ &$0.74$ &11.5 &5.5 &$2.54$
\\
& & & & & & & & & & \\
J1415--66 & \ldots & 26.3 & 6.5 &\ldots &$-48.8$ & $-4.8$ &$14$ &$1.2$ &\ldots &\ldots &\ldots 
\\
J1424--69 & \ldots & 13.0 & 3.6 &\ldots &$-49.2$ & $-8.5$ &$6.3$ &$0.94$ &\ldots &\ldots &\ldots 
\\
J1517--4356 & 5.6 & 11.8 & 6.1 &14.4 &$-31.1$ & $11.5$ &$4.4$ &$0.88$ &47.8 &0.38 &$30.9$
\\
J1528--4109 & 3.7 & 19.8 & 6.4 &12.3 &$-27.9$ & $12.7$ &$6.0$ &$1.3$ &21.1 &0.46 &$107$
\\
J1531--4012 & 5.3 & 15.4 & 11.1 &\ldots &$-26.9$ & $13.1$ &$>7.8$ &$>1.8$ &58.7 &0.19 &$83.6$
\\
& & & & & & & & & & \\
J1535--4114 & 3.8 & 95.0 & 12.5 &18.3 &$-26.8$ & $11.8$ &$2.8$ &$0.57$ &1.68 &1.3 &$1981$
\\
J1536--44 & \ldots & 15.8 & \ldots &\ldots &$-28.5$ & $9.3$ &$5.3$ &$0.85$ &\ldots &\ldots &\ldots 
\\
J1537--49 & \ldots & 14.5 & 26.3 &\ldots &$-31.3$ & $5.2$ &$1.7$ &$0.16$ &\ldots &\ldots &\ldots 
\\
J1540--63 & \ldots & 25.3 & 12.4 &\ldots &$-39.4$ & $-6.5$ &$7.5$ &$0.84$ &\ldots &\ldots &\ldots 
\\
J1603--3539 & 4.1 & 13.3 & 30.4 &\ldots &$-18.8$ & $12.5$ &$3.8$ &$0.83$ &18.1 &0.13 &$1716$
\\
& & & & & & & & & & \\
J1617--4216 & 5.1 & 10.3 & 3.7 &9.5 &$-21.5$ & $5.9$ &$6.3$ &$0.65$ &3.00 &8.0 &$17.8$
\\
J1641--2347 & 3.7 & 85.1 & 16.2 &29.7 &$-4.2$ & $14.7$ &$1.3$ &$0.34$ &421 &0.21 &$1.25$
\\
J1649--5553 & 7.3 & 13.5 & 39.9 &\ldots &$-27.9$ & $-7.1$ &$14$ &$1.7$ &5.73 &1.0 &$290$
\\
J1655--3048 & \ldots & 32.3 & 57.0 &72.9 &$-7.8$ & $7.9$ &$8.6$ &$1.2$ &235 &0.14 &$9.02$
\\
J1701--3130 & 5.9 & 26.8 & 11.6 &23.5 &$-7.5$ & $6.4$ &$4.5$ &$0.50$ &82.5 &0.13 &$89.3$
\\
\end{tabular}
\end{minipage}
\end{table*}
\begin{table*}
\begin{minipage}{15cm}
\contcaption{}
\begin{tabular}{llllllllllll}
Name & $\Delta$ pos & S/N& $w_{50}$ & $w_{10}$ & $l$ & $b$ & $d$ & $|z|$ & $\tau_c$ & $B$ & $\dot{E}$\\
 &   (\arcmin) & & (\degr) & (\degr) & (\degr) & (\degr) & (kpc) & (kpc)  & (Myr)& $(10^{12}$ G) & ($10^{30}$~erg~s$^{-1}$)\\
\hline
& & & & & & & & & & \\
J1706--61 & \ldots & 21.6 & 2.8 &\ldots &$-30.8$ & $-12.1$ &$3.8$ &$0.80$ &\ldots &\ldots &\ldots 
\\
J1712--2715 & 1.1 & 41.7 & 46.5 &90.2 &$-2.7$ & $7.1$ &$3.1$ &$0.38$ &3.16 &0.58 &$3033$
\\
J1717--5800 & 5.0 & 10.9 & 23.5 &\ldots &$-27.3$ & $-11.5$ &$>8.8$ &$>1.8$ &26.1 &0.25 &$232$
\\
J1721--1939 & 4.6 & 9.7 & 16.9 &\ldots &$4.9$ & $9.6$ &$4.7$ &$0.78$ &49.9 &0.23 &$76.8$
\\
J1739--1313 & 3.3 & 22.6 & 2.4 &5.3 &$12.8$ & $9.3$ &$2.0$ &$0.33$ &236 &0.32 &$1.80$
\\
& & & & & & & & & & \\
J1741--2019 & 6.0 & 59.6 & 11.1 &13.8 &$6.8$ & $5.4$ &$2.0$ &$0.19$ &3.80 &8.1 &$10.8$
\\
J1742--4616 & 3.1 & 26.3 & 21.8 &33.2 &$-15.2$ & $-8.5$ &$5.0$ &$0.74$ &193 &0.12 &$19.0$
\\
J1743--4212 & 5.2 & 19.6 & 8.6 &18.8 &$-11.6$ & $-6.5$ &$4.7$ &$0.53$ &6.19 &0.50 &$1078$
\\
J1744--1610 & \ldots & 12.8 & 7.4 &10.2 &$10.8$ & $6.9$ &$2.0$ &$0.24$ &11.7 &2.1 &$17.3$
\\
J1745--0129 & 5.4 & 12.2 & 2.4 &8.3 &$23.8$ & $14.0$ &$>7.3$ &$>1.8$ &26.3 &0.82 &$21.8$
\\
& & & & & & & & & & \\
J1802+0128 & 1.9 & 12.5 & 4.1 &\ldots &$28.6$ & $11.6$ &$>8.8$ &$>1.8$ &4.16 &1.1 &$489$
\\
J1805--0619 & 6.0 & 11.1 & 12.7 &23.0 &$22.0$ & $7.2$ &$6.7$ &$0.84$ &7.43 &0.67 &$407$
\\
J1806+10 & \ldots & 66.3 & 6.7 &\ldots &$37.3$ & $14.6$ &$4.0$ &$1.0$ &\ldots &\ldots &\ldots 
\\
J1808--3249 & 2.1 & 40.8 & 13.8 &20.7 &$-1.0$ & $-6.1$ &$5.1$ &$0.54$ &0.82 &1.6 &$5727$
\\
J1809--0743 & 1.3 & 13.0 & 13.8 &\ldots &$21.2$ & $5.7$ &$>18$ &$>1.8$ &32.7 &0.22 &$194$
\\
& & & & & & & & & & \\
J1811--0154 & 3.8 & 21.8 & 8.2 &77.8 &$26.6$ & $8.0$ &$9.6$ &$1.3$ &9.12 &1.2 &$80.2$
\\
J1819+1305 & 5.1 & 14.3 & 21.4 &\ldots &$41.2$ & $12.8$ &$4.4$ &$0.98$ &45.0 &0.64 &$12.4$
\\
J1824--25 & \ldots & 15.6 & 9.3 &\ldots &$7.1$ & $-5.9$ &$5.2$ &$0.53$ &\ldots &\ldots &\ldots 
\\
J1832--28 & \ldots & 18.0 & 21.3 &\ldots &$5.1$ & $-8.9$ &$6.4$ &$1.00$ &\ldots &\ldots &\ldots 
\\
J1837+1221 & 4.9 & 16.1 & 3.1 &\ldots &$42.4$ & $8.7$ &$6.1$ &$0.93$ &5.02 &3.5 &$32.3$
\\
& & & & & & & & & & \\
J1837--1837 & 4.8 & 11.0 & 5.8 &11.6 &$14.8$ & $-5.5$ &$3.0$ &$0.29$ &1.78 &1.9 &$918$
\\
J1842+1332 & 3.5 & 33.1 & 72.5 &\ldots &$44.1$ & $8.1$ &$5.9$ &$0.83$ &32.6 &0.33 &$86.2$
\\
J1848+12 & \ldots & 17.8 & 51.0 &\ldots &$44.1$ & $6.5$ &$7.5$ &$0.84$ &\ldots &\ldots &\ldots 
\\
J1855--0941 & 2.9 & 19.1 & 26.1 &\ldots &$24.7$ & $-5.2$ &$4.9$ &$0.45$ &22.8 &0.29 &$230$
\\
J1857--1027 & 0.9 & 71.4 & 14.6 &20.5 &$24.3$ & $-6.1$ &$3.6$ &$0.38$ &5.54 &6.3 &$8.31$
\\
& & & & & & & & & & \\
J1901--1740 & 5.5 & 131.0 & 6.6 &22.1 &$18.1$ & $-10.1$ &$1.3$ &$0.22$ &37.7 &1.3 &$4.34$
\\
J1919+0134 & 5.0 & 34.2 & 10.5 &18.2 &$37.6$ & $-5.6$ &$10$ &$0.99$ &43.1 &0.98 &$5.64$
\\
J1943+0609 & 2.5 & 30.1 & 8.9 &17.7 &$44.5$ & $-8.6$ &$3.9$ &$0.58$ &15.2 &0.46 &$207$
\\
J1947+0915 & 6.1 & 12.7 & 7.5 &\ldots &$47.7$ & $-8.1$ &$5.8$ &$0.81$ &49.1 &0.85 &$5.81$
\\
J1956+0838 & 5.8 & 15.0 & 15.4 &\ldots &$48.3$ & $-10.3$ &$4.3$ &$0.77$ &21.9 &0.26 &$309$
\\
& & & & & & & & & & \\
J2007+0809 & 5.2 & 14.4 & 92.8 &\ldots &$49.2$ & $-12.8$ &$3.4$ &$0.76$ &37.7 &0.21 &$156$
\\
\end{tabular}
\end{minipage}
\end{table*}

\begin{figure}
\centerline{\psfig{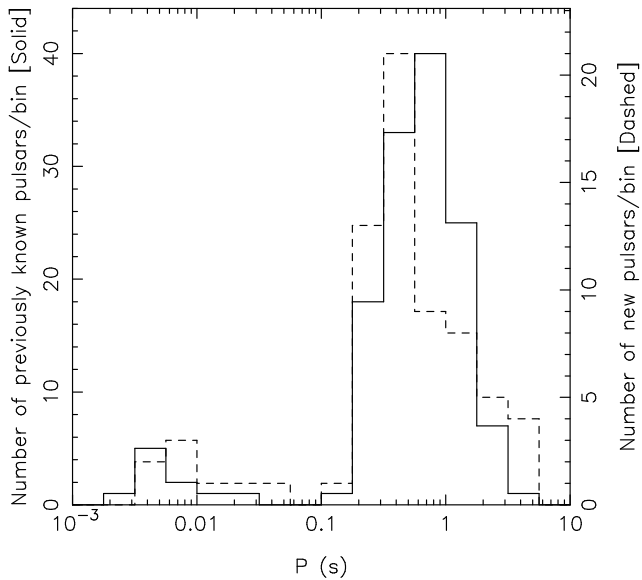}}
\caption{Histograms depicting distribution in pulse period of
new pulsars (dashed line) and previously known pulsars in the survey
region (solid line).}
\label{fig:phist}
\end{figure}
\begin{figure}
\centerline{\psfig{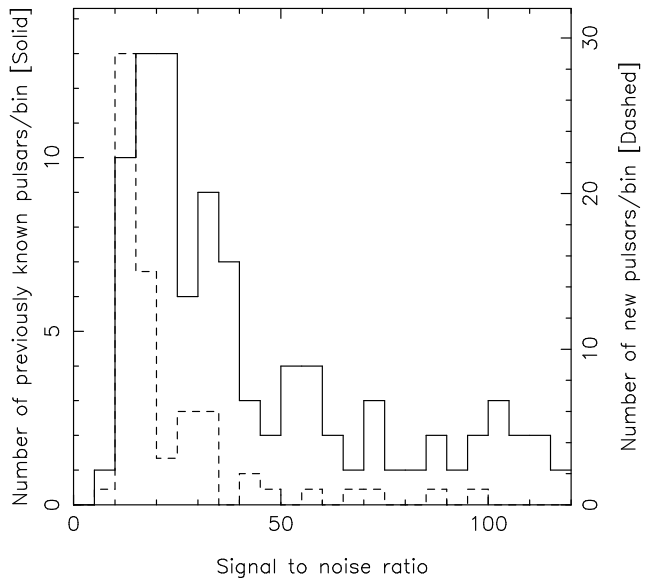}}
\caption{Histograms depicting distribution in maximum detected signal
to noise ratio (S/N) of new pulsars (dashed line) and previously known
pulsars in the survey region (solid line). One new and 8 known pulsars
had S/Ns greater than 120 and are not included in this plot for
clarity in low S/N bins.}
\label{fig:snhist}
\end{figure}
\begin{figure}
\centerline{\psfig{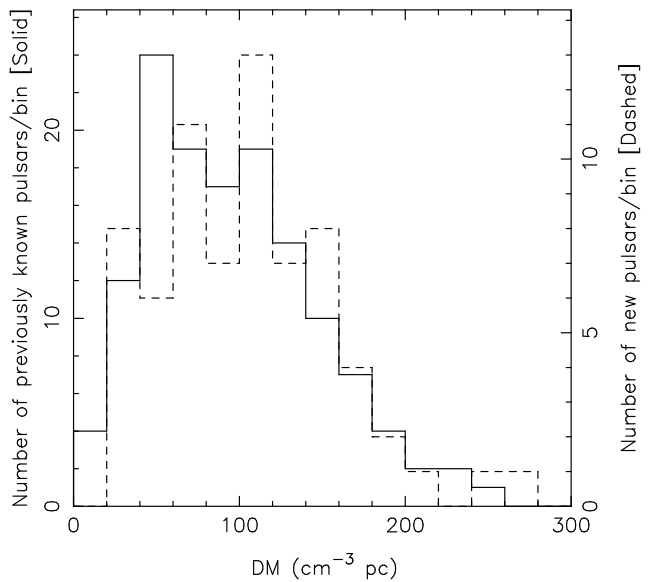}}
\caption{Histograms depicting distribution in dispersion measure of
new pulsars (dashed line) and previously known pulsars in the survey
region (solid line).}
\label{fig:dmhist}
\end{figure}
\begin{figure}
\centerline{\psfig{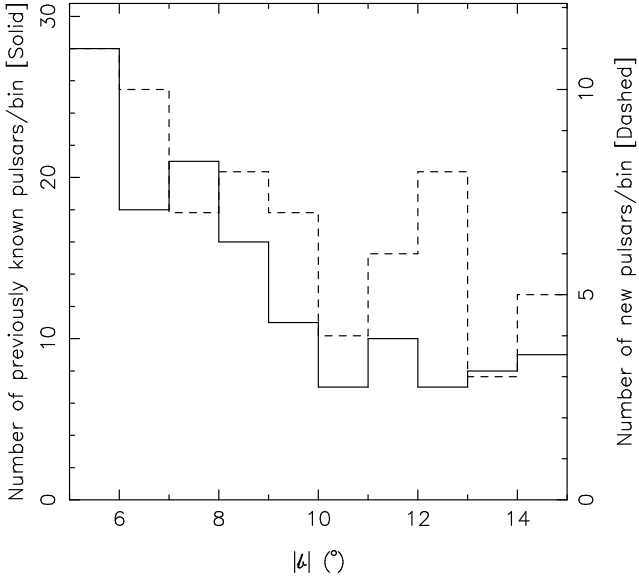}}
\caption{Histograms depicting distribution in angular displacement
from the Galactic plane of new pulsars (dashed line) and previously
known pulsars in the survey region (solid line).}
\label{fig:bhist}
\end{figure}

A total of 101 previously known pulsars were re-detected in the
survey. Due to the large angular extent of each of the tessellated
units of four pointings, the actual survey regions are significantly
non-rectangular in $l$ and $b$ (see Figure \ref{fig:coverage}) and
three of these pulsars actually lie outside the nominal survey
region. There were 135 previously discovered pulsars in the declared
region, leaving 37 undetected in the survey. Tables \ref{tab:knowndet}
and \ref{tab:knownnondet} list the detected and undetected known
pulsars, including for each pulsar the period, dispersion measure,
flux density at 1400 MHz (where known), position in Galactic
coordinates, angular offset from the centre of the beam in which it
was detected (where known, or the nearest processed beam for
undetected pulsars), and the signal to noise ratio of detections. 
In most cases the values for period, DM and flux density derive from
the works of \scite{tml93}, \scite{lylg95} or \scite{dsb+98}.

From the tabulated position offsets to the nearest processed beams and
given the fact that the centres of adjacent beams of the grid are
spaced 14 arcminutes apart on the sky, it is clear that the reason for
many of the non-detections was incomplete sky coverage. Twelve
undetected pulsars lay greater than 10 arcminutes from the nearest
beam, leading to an alternative completeness metric of $1-12/135=91$
per cent, (or $1-9/135=93$ per cent if the loss is offset by the three
known pulsars detected outside the survey region). Of the remaining 25
non-detections, 11 have published flux densities near 1400 MHz and are
plotted along with the sensitivity curves of Figure
\ref{fig:sensitivity}. Allowing for scintillation, all are compatible
with having flux densities below the sensitivity limit.  We therefore
expect that most of those pulsars lacking flux density measurements
were also below the detection threshold at the time of observation,
and conclude that the search procedure was adequate and robust in its
rejection of interference without significant loss of pulsars.
Possible exceptions to this statement are PSR B1556--44 which was
detected at 17 times its true spin frequency (probably due to
proximity in period to a persistent 256-ms interference signal), and
perhaps PSR J1130--6807 which has a similar pulse period but being of
unknown flux density may have simply fallen below the sensitivity
limit. The discovery of PSR J1712--2715 (below) with $P=255.4$ ms
indicates that rough proximity to interference signals is not always
problematic. The millisecond pulsar (MSP) fraction of the undetected
pulsars is high, however all except J1911--1114 \cite{llb+96} were
discovered in deep directed searches of globular clusters
\cite{lbm+87,bbl+94,dlm+01} which found mainly millisecond
pulsars. Figure \ref{fig:fluxhist} shows a histogram of flux densities
for previously known pulsars of published flux density with processed
beams centred less than 10$\arcmin$ away, with the distribution of
non-detections hashed. As expected from the sensitivity curves shown
in Figure \ref{fig:sensitivity}, the survey was sensitive to most
pulsars brighter than 1~mJy, insensitive to pulsars with $S <
0.1$~mJy, and recorded a mixture of detections and non-detections in
the remaining range due to scintillation and the distribution of pulse
widths.

Examination of the detection parameters of previously known pulsars
reveals some discrepancies with previously published results. From
inspection of the position offsets of newly discovered pulsars (below,
Table \ref{tab:newparams2}), we estimate a position uncertainty of
7\arcmin\ for detections in this survey, consistent with the beam
spacing.  Of those detected pulsars with published positions greater
than 10\arcmin\ from the beam centre, two were not
detected in pointings made closer to the published position. The
(B1950) declinations reported for PSRs B1232--55 and B1359--51 of
$-55\degr00(10)\arcmin$ and $-51\degr10(15)\arcmin$ \cite{mlt+78}
respectively are inconsistent with the B1950 declinations of
$-54\degr40(5)\arcmin$ and $-50\degr06(5)\arcmin$ of the detections
made in this survey. The published right ascension values have much
smaller uncertainties (due to the shape of the beam of the Molonglo
telescope with which they were discovered), and are consistent with
our detections.

As was
spectacularly illustrated with PSR J2144--3933 \cite{ymj99}, pulsar
surveys in the past have been prone to detecting pulsars with an
apparent pulse frequency an integer multiple of the true frequency.
We expect that this is due to the flattening of the power spectrum
with a boxcar filter, the presence of interference signals of similar
period, the dominance of odd harmonics in the pulse shape, or the
exclusion of the fundamental as being below a cutoff frequency (as was
the case for J2144--3933). We found that PSRs J1403--7646, J1817--3837
and J1901--0906 \cite{lml+98} actually possess periods a factor of two
greater than the published values. Conversely, pulsars J1036--4926,
B1110--69, B1524--39, B1556--44, B1706--16, B1709--15, B1717--16,
B1718--19 and B1848+12 were erroneously re-detected with shorter pulse
periods. All but one of these results were made in the early stages of
processing before interference rejection had been fine-tuned to avoid
accidental filtering of pulsar harmonics near interference
signals. After correction for such errors and with the exception of
the three pulsars listed above, all other detections were made with
pulse periods consistent with previously published parameters.

\subsection{Newly Discovered Pulsars}
\label{sec:mbsurv.newpsrs}

The initial viewing of the survey results and subsequent confirmation
observations resulted in the discovery of 58 new pulsars, 8 of which
possess short spin periods and small period derivatives indicative of
recycling and have been reported elsewhere \cite{eb01a,eb01b}. The
final careful review of candidates produced a further 11 slow
pulsars. Data from all pulsars were folded at twice and three times
the discovery period to detect any errors of the kind described in the
previous section. Pulsars J1055--6905, J1517--4356, J1802+0128 and
J1808--3249 were all initially discovered at half the true spin
period, however J1517--4356 and J1808--3249 were later detected at the
correct period in subsequent survey observations, with higher signal
to noise ratio. Pulse profiles for the 61 slow pulsars discovered in
the survey are presented in Figure \ref{fig:profs}. For those pulsars
with a timing solution the profile arises from the
summation of numerous good observations, whilst for others the profile
from the single best observation made to date is provided. It should
be noted that the baselines of some profiles are corrupted due to the
response of the filterbank/digitiser low pass filter.

Timing solutions have been obtained for 48 slow pulsars. The data for
each pulsar span 490--830 days, depending on when the candidate
re-observation was made. Since all pulsars have been timed for well
over a year, we were able to accurately measure positions and period
derivatives, in addition to making improved
measurements of periods and dispersion measures.  We have not seen
obvious timing noise in any pulsar, however it is possible that such
noise is present and is absorbed by deviations from the ``true''
values of the fitted parameters. Dispersion measures were fitted for
by the use of TOAs derived from sub-divisions of the observing band,
or where available, 660~MHz observations. In the latter case the
signal to noise ratio available in a reasonable integration time was
poor and the use of an independent template profile was eschewed in
favour of the superior 20-cm template.

Table \ref{tab:newparams1} presents the basic parameters of the slow
pulsars discovered in this survey, derived from timing measurements
when available.  Values in parentheses denote uncertainties in the
last quoted digit and represent twice the formal uncertainty produced
by TEMPO. In the case of pulsars lacking a timing solution we use
twice the errors derived from the fine $P$-DM search, or for
positions, errors corresponding to a randomly oriented offset of 7
arcminutes (i.e. $7\arcmin/\sqrt{2} \simeq 5\arcmin$ in each of
$\alpha$ and $\delta$).  PSR J1802+0128 was timed for 493 days before
it was discovered that the true spin period was twice the assumed
value. Values presented for $P$ and $\dot{P}$ are twice the values
derived by this timing analysis and retain their prior relative
errors. Due to the effects of the evolution of profile morphology with
frequency and the use of the 20~cm as template for timing 50~cm
observations, the formal errors presented for the dispersion measure
may have been underestimated. Pulsar names are assigned from their
equatorial coordinates in the J$2000.0$ equinox, with four digits of
declination for those with accurate positions from timing solutions,
and two digits for those without. The latter names are provisional and
will be altered when accurate timing positions become available. Table
\ref{tab:newparams2} lists additional parameters for the newly
discovered pulsars, including the best S/N of the discovery
observation(s) and the pulsar's position offset from the centre of the
beam, the width of the pulse profile at 10 and 50 per cent of peak
intensity, its position in Galactic coordinates, its distance and
$z$-displacement from the Galactic plane under the model of
\scite{tc93} (accurate to 30 per cent on average), and inferred
parameters concerning the pulsar spin-down. These assume magnetic
dipole spin-down and comprise the characteristic age ($\tau_c \equiv
P/2\dot{P}$), surface magnetic field strength ($B = 3.2\times10^{19}
$~G~s$^{-1/2}\sqrt{P\dot{P}}$) and spin-down power ($\dot{E} =
4\pi^2I\dot{P}P^{-3}$, assuming $I=10^{45}$~g~cm$^2$ for the moment of
inertia of the neutron star). Both tables are accessible on the
internet in machine-readable format at the Swinburne Pulsar Group home
page\footnote{http://www.astronomy.swin.edu.au/pulsar}. For several
pulsars the dispersion measure is higher than that allowed in the
given direction under the model of \scite{tc93}, and the values
presented are given as lower limits.  Such findings are not uncommon
(e.g. \pcite{cn95,dsb+98}), and indicate that the model probably
underestimates the scale height of the Galactic electron distribution.

The spin parameters of the newly discovered systems are similar to
those of pulsars previously known in the search region.  Figure
\ref{fig:ppdot} shows the distribution in period and period derivative
of new pulsars with timing solutions and of previously known pulsars
inside and outside the survey region. Both new and previously known
slow pulsars in the region tend to have longer inferred characteristic
ages than those outside the survey region, by simple virtue of the
fact that pulsars are born near the Galactic plane and typically take
several Myr to reach a $z$-height corresponding to $|b| > 5\degr$ (for
typical distances of several kpc). Figure \ref{fig:phist} shows
histograms of pulse period for the new and previously known
population. It appears that this survey has uncovered a higher
fraction of pulsars in the period range of 6--50~ms, however this
effect is not highly significant : a Kolmogorov-Smirnov (K-S) test on
the distributions for $P < 100$~ms yields a 46 per cent probability of
the two samples arising from the same parent distribution. When binary
evolution considerations are taken into account and the new sample
restricted to the four MSPs with probable helium companions the
significance rises to 91 per cent, however due to the small sample
size this result must be treated with caution \cite{eb01b}.

The discrepancy in the case of
the slow pulsar population is more significant, with a K-S test
indicating a different distribution at the 98 per cent level. The
reason for the discrepancy, which is mainly seen as a deficit of
pulsars with $P\simeq 1$~s, is not well understood. The problem
remains (at 94 per cent significance) when the new sample is compared
only to those previously known pulsars re-detected in this survey,
confirming that the survey was able to detect pulsars in the period
range in question. However, if there existed (by an unknown mechanism)
a reduced sensitivity (or higher effective threshold signal to noise
ratio) around $P\simeq 1$~s, one might expect the period distribution
of the new pulsars to be more strongly affected than that of the
previously known pulsars since the new population is on average of
lower flux density. We note that the rejection of pulsars as
mis-categorized interference signals cannot explain this result, since
in this case one would expect an equal rate of rejections of new and
previously known pulsars independent of signal to noise ratio or flux
density.  Comparison of the (yet to be measured) flux densities of new
pulsars in and out of the depleted period range will help in
evaluating the effective sensitivity of the survey as a function of
period.

Figure \ref{fig:snhist} shows the distribution of signal to noise
ratios in the best detections of new and previously known pulsars. As
one would expect, most of the pulsars with high signal to noise ratios
were detected in earlier surveys. A few of the new pulsars however
were very strong and may have been missed in earlier surveys due to
scintillation or due to an intrinsically flat spectrum. It is apparent
from the histogram and from inspection of Tables \ref{tab:knowndet}
and \ref{tab:newparams2} that the threshold signal to noise ratio for
this survey is approximately 10, in contrast to the value of $8.0$
commonly used in previous surveys in assessing sensitivity. Numerous
promising candidates with signal to noise ratios in the range of 8--10
were subjected to re-observation however only one was re-detected in
such observations despite using longer integration times, probably
because they actually arose from interference or by random chance
(given the size and dimensionality of the search space).

\begin{figure*}
\centerline{\psfig{figure=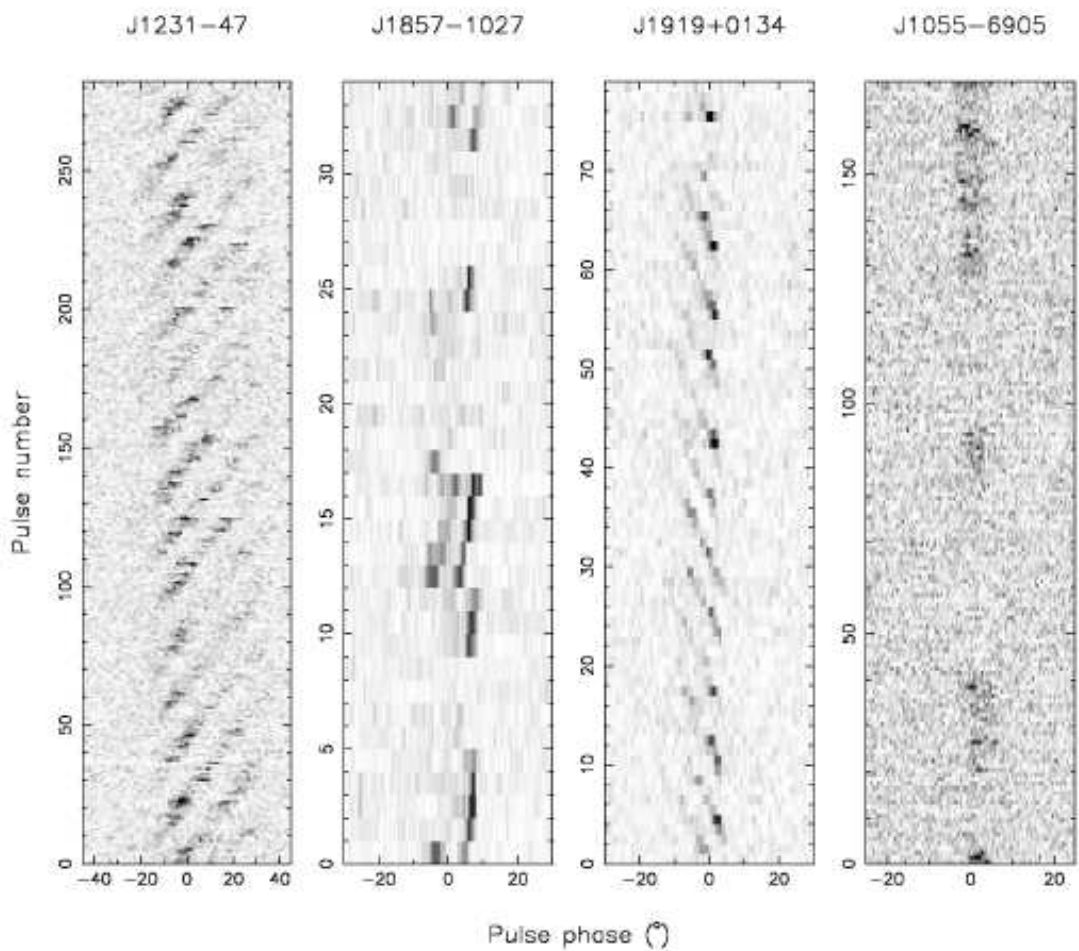,width=15cm}}
\caption{Greyscale plots of detected flux density as a function of
pulse phase and pulse number for four newly discovered pulsars. Each
row represents a single pulse.}
\label{fig:drifters}
\end{figure*}

As noted in the introduction, high frequency surveys are also expected
to sample a different area of the pulsar spectral index distribution,
compared to low frequency surveys.  The sensitivity of the present
work to a pulsar with a spectral index of $-1.7$ (typical of those
discovered at 70~cm; \pcite{tbms98}) is comparable to that of the
Parkes Southern Pulsar Survey \cite{mld+96}, the most sensitive
previous 70~cm search to cover a large region of the area observed by
the present study.  One therefore expects the distances of the newly
discovered pulsars to be comparable to the previously known population
in the region, and as shown in Figure \ref{fig:dmhist}, this is indeed
the case.  The bulk of newly discovered pulsars had S/Ns $\la 30$,
suggesting (in conjunction with their non-detection in the 70~cm
survey) spectral indices of up to $-0.7$. Eleven newly discovered pulsars
are visible from the 305-m Arecibo telescope and were presumably
within the search area of previous surveys conducted there
\cite{fcwa95,cnst96,rtj+96,lzb+00}, which were typically sensitive to
($\delta=0.1$) pulsars brighter than $\sim1$~mJy at 70~cm. Their
non-detection in the Arecibo surveys suggests $\alpha \ga -1.1$.
Several pulsars were discovered at high S/N, which could be indicative
of positive spectral indices (for example, J1806+10 in Arecibo
territory, yielding $\alpha \simeq 2.6$), however the non-detections
may well be the result of incomplete surveys or interstellar
scintillation.

Accurate characterization of the spectral indices of the newly
discovered pulsars must await calibrated multi-frequency flux density
measurements, however we note in passing that the detection of these
pulsars at 660~MHz required significantly more integration time than
expected for pulsars of average spectral index.  As indicated by the
lack of any enhanced preference for low Galactic latitudes in the
newly detected sample compared to the previously known pulsars (Figure
\ref{fig:bhist}), reduced scatter-broadening at 20-cm in general does
not appear to have been a significant factor in the discovery of the
new pulsars.

\subsection{Individual Pulsars of Interest}
\label{sec:mbsurv:individual}
As previously reported \cite{eb01b}, the pulse profile of PSR
J1410--7404 is exceedingly narrow and contradicts the pulse width --
period relation of \scite{ran90}. Since all major
contradictions in the past have been from apparently recycled pulsars,
it is conceivable that J1410--7404 is also recycled, a hypothesis
supported by the small magnetic field strength inferred from
timing observations. The newly discovered pulsar J1706--61 also has a
measured profile width seemingly in disagreement with \scite{ran90},
in this case being 70 per cent of the predicted minimum width. However, we
caution that this measurement derives from a single observation of
moderate signal-to-noise ratio (as visible in Figure \ref{fig:profs})
and that a more definite conclusion awaits the availability of an
extended data set from ongoing timing observations. Should the
improved profile maintain the narrow width derived here, the magnetic
field strength and characteristic age derived from timing measurements
will be of great utility in evaluating the recycling hypothesis for
PSRs J1410--7404 and J1706--61.

As expected from the relatively old age of the sample detected here,
numerous pulsars appear to exhibit noticeable pulse nulling
(e.g. \pcite{rit76}). Several pulsars also show drifting sub-pulses
(e.g. \pcite{bac73}) of varying degrees of regularity. Figure
\ref{fig:drifters} depicts the instantaneous flux density of four newly
discovered
pulsars as a function of pulse phase and pulse number. It is apparent
that the emission of PSRs J1231--47 and J1919+0134 is strongly
modulated by sub-pulses showing very regular drift. PSR J1857--1027
also appears to show drifting sub-pulses in this and other
observations, although as a result of the short duration of
observations made for timing analysis,
combined with the long pulse period ($\sim 3.7$~s), the number of
pulses recorded and hence the conspicuousness of this effect in
archival observations is reduced. Detailed analysis of these pulsars
will appear in a forthcoming paper (Ord \etal\ in preparation).

 Pulse nulling appears to occur on a wide range of time scales in the
detected pulsars. The data for PSR J1231--47 in Figure
\ref{fig:drifters} were recorded in a second attempt at confirmation
of this pulsar, resulting in a signal to noise ratio of 82 in an
observation of 530~s. Several subsequent observations of between 900
and 4000~s succeeded in producing only one further weak detection of
the pulsar. PSRs J1857--1027 and J1055--6905 appear to null on more
typical timescales of $\sim$1--50 pulses, as shown in Figure
\ref{fig:drifters}. Future observations of the candidate nulling
pulsars from our sample will enable analysis similar to that of
\scite{rit76}, which in turn will help decide whether the observed
flux density variations arise due to nulling or simply as a result of
interstellar scintillation.

\section{Conclusions}
We have conducted a survey of intermediate latitudes of the southern
Galaxy for pulsars at $\sim$1400~MHz. The new 13-beam 21-cm receiver
of the Parkes radio telescope was used to rapidly cover to moderate
depth a large region of sky flanking the area of the deeper ongoing
Galactic plane survey \cite{lcm+00,clm+00}. The interference
environment was formidable, however development of a comprehensive
scheme for the rejection of pulsar candidates arising from
interference enabled the realisation of the full expected survey
sensitivity of approximately $0.5$~mJy for slow and most millisecond
pulsars. The survey was highly successful, detecting 170 pulsars of
which 69 were previously unknown, in a relatively short observing
campaign. The new discoveries are not significantly more distant than
the previously known population in this region of sky, indicating that
the success of the survey is attributable to its sampling of a
different portion of the broad distribution of pulsar spectral
indices. The detected sample, in combination with those of the
Galactic plane and high-latitude surveys (when complete), will prove
invaluable for population modelling due to the use of a common
observing system to cover a large area of sky at high radio frequency.

Among the most interesting new objects are two recycled
pulsars with massive white dwarf companions \cite{eb01a}, four with
probable low-mass He dwarf companions, two isolated millisecond
pulsars, and one `slow' pulsar with a very narrow pulse profile and
small period derivative, suggestive of recycling in a scenario similar
to those of the known double neutron star systems \cite{eb01b}. As
expected from the large Galactic $z$-height of much of the survey
volume, the detected population of slow pulsars was relatively old and
as such exhibited a high fraction of pulsars showing nulling and
sub-pulse modulation. Two pulsars show very regular drifting
sub-pulses and are analysed in detail elsewhere (Ord \etal\ in
preparation).

\section*{acknowledgments}
We thank the members of the Galactic plane Parkes multibeam pulsar
survey collaboration for the use of equipment built for that survey,
and for the exchange of observing time to improve the regularity of
timing observations. The free exchange of software and information
between the Galactic plane collaboration and the authors was a
pleasant and useful arrangement.  We are grateful for the high level
of support provided by the staff of the CSIRO/ATNF Parkes radio
telescope in the event of system failures. We thank the referee for
detailed comments and suggestions which improved the manuscript.  RTE
acknowledges the support of an Australian Postgraduate Award. MB is an
ARC Senior Research Fellow, and this research was supported by the ARC
Large Grants Scheme.


\end{document}